# On Asymmetry of Magnetic Helicity in Emerging Active Regions: High Resolution Observations

Lirong Tian[1], Pascal Démoulin[2], David Alexander[1], and Chunming Zhu[1]

[1]*Department of Physics and Astronomy, Rice University, 6100 Main Street, Houston, TX 77251-1892, USA*

[2] *Observatoire de Paris, LESIA UMR 8109 (CNRS), 5 place Jules Janssen, 92190 Meudon, France*

## ABSTRACT

We employ the DAVE (differential affine velocity estimator, Schuck 2005; 2006) tracking technique on a time series of MDI/1m high spatial resolution line-of-sight magnetograms to measure the photospheric flow velocity for three newly emerging bipolar active regions. We separately calculate the magnetic helicity injection rate of the leading and following polarities to confirm or refute the magnetic helicity asymmetry, found by Tian & Alexander (2009) using MDI/96m low spatial resolution magnetograms. Our results demonstrate that the magnetic helicity asymmetry is robust being present in the three active regions studied, two of which have an observed balance of the magnetic flux. The magnetic helicity injection rate measured is found to depend little on the window size selected, but does depend on the time interval used between the two successive magnetograms tracked. It is found that the measurement of the magnetic helicity injection rate performs well for a window size between 12×10 and 18×15 pixels, and at a time interval $\Delta t$=10 minutes. Moreover, the short-lived magnetic structures, 10−60 minutes, are found to contribute 30−50% of the magnetic helicity injection rate. Comparing with the results calculated by MDI/96m data, we find that the MDI/96m data, in general, can outline the main trend of the magnetic properties, but they significantly underestimate the magnetic flux in strong field region and are not appropriate for quantitative tracking studies, so provide a poor estimate of the amount of magnetic helicity injected into the corona.

## 1. Introduction

Coronal mass ejections (CMEs) are energetic expulsions of plasma from the solar corona that are driven by the release of a large amount of magnetic energy and magnetic helicity, which significantly affect interplanetary and geomagnetic space. There is very strong evidence from many studies of the existence of highly twisted and sheared magnetic field structures associated with CMEs and large flares, such as $H_\alpha$ fibrils in filaments and coronal X-ray sigmoids. The coronal magnetic field stores progressively large amounts of magnetic energy and helicity before becoming unstable and producing a flare and/or a CME. A large fraction of the magnetic energy and helicity is believed to be injected from the convection zone through the photosphere into the corona during the emergence of an active region (Leka et al. 2006; Jeong & Chae, 2007; Tian & Alexander, 2008; see also a review of Démoulin & Pariat 2009; van Driel-Gesztelyi & Cullane 2009; and references therein).

Magnetic helicity is an important global quantity of magnetohydrodynamic (MHD) theory (e.g. Berger & Field, 1984). Its conservation property is a powerful tool to relate a solar coronal mass ejection to its interplanetary counterpart, independently of the amount of internal magnetic reconnection taking place. This conservation property also provides constrains to the solar dynamo since the amount of magnetic helicity which is observed to cross the photosphere into the corona should have been produced below by the solar dynamo (see reviews of Démoulin 2007; Démoulin & Pariat 2009; Fan 2009; van Driel-Gesztelyi & Culhane 2009; and references therein).

Unlike the magnetic helicity of closed magnetic fields, the helicity of open magnetic fields may change with time as magnetic helicity is transported across the boundaries. Therefore the magnetic helicity of the coronal volume of an active region is a function of time. In order to follow the evolution of the magnetic helicity content of the coronal volume enclosing an active region, we can determine the magnetic helicity flux at the photospheric level (Berger & Field 1984; see also Démoulin & Berger 2003),

$$\begin{aligned} \frac{dH}{dt} &= -2\int_S [(\mathbf{A}_p \cdot \mathbf{v})\mathbf{B} - (\mathbf{A}_p \cdot \mathbf{B})\mathbf{v}] \cdot \mathrm{d}\mathbf{S} \\ &= -2\int_S (\mathbf{A}_p \cdot \mathbf{v}_h) B_n \, \mathrm{d}S + 2\int_S (\mathbf{A}_p \cdot \mathbf{B}_h) v_n \, \mathrm{d}S \\ &= \left.\frac{dH}{dt}\right|_h + \left.\frac{dH}{dt}\right|_n . \end{aligned} \quad (1)$$

The first term on the right hand represents the magnetic helicity contribution of horizontal mass motions parallel to the photosphere; the second term represents the helicity contribution of the emerging twisted/sheared magnetic field. $\mathbf{v}_h$ and $\mathbf{B}_h$ are horizontal (transverse) components of mass flow velocity and magnetic field respectively, while $v_n$ and $B_n$ are the respective vertical (normal) components. $\mathbf{A}_p$ is the vector potential of the potential magnetic field computed from the photospheric $B_n$ distribution (with $A_n = 0$ at the photosphere and $\nabla \cdot \mathbf{A} = 0$ in the coronal volume).

In order to determine observationally the amount of magnetic helicity injected into the corona of an active region using Equation (1), we must have acceptable time series of vector magnetograms ($\mathbf{B}_h$ and $B_n$) and detailed observations of photospheric mass flow velocity ($\mathbf{v}_h$ and $v_n$). The vertical component of the velocity ($v_n$) may be approximately determined from Doppler velocities near disk center when available. In general, the horizontal component ($\mathbf{v}_h$) is determined by employing one of many tracking techniques developed in recent years, such as the induction method (IM, Kusano et al. 2002, 2004), inductive local correlation tracking (ILCT, Welsch et al. 2004), minimum energy fit (MEF, Longcope 2004), and differential affine velocity estimator for vector magnetograms (DAVE4VM, Schuck 2008), based on the time series of the vector magnetograms. The performance of the methods have been compared, and the results are reported in papers of Welsch et al. (2007), Chae & Sakurai (2008), and Schuck (2008). The DAVE4VM technique is designed to extend the DAVE approach to handle vector magnetic fields (Schuck 2008).

However, the vector magnetograms are presently not good enough for the study of the magnetic helicity injection. The Solar Optical Telescope (SOT) on Hinode has been providing vector data of active regions since it started taking observations by the end of 2006. However, data download problems, mainly related to telemetry issues, have meant that the vector magnetograms are not recorded with a sufficiently high time cadence for an accurate derivation of the velocity flow field and are therefore have not proved useful for this study. The Helioseismic and Magnetic Imager (HMI) on Solar Dynamics Observatory (SDO), launched on February, 2010, will provide many vector magnetograms with high spatial resolution, and should significantly enhance flux emergence studies of the sort reported here. However, a number of key issues will remain, namely the challenging of 180° ambiguity of the orientation of the transverse field components and the inter-calibration between the transverse and vertical components. Moreover, the transverse field components are much noisier than the line-of-sight field component (about 100G versus 10G), thus some magnetic helicity contribution from shearing and twisting magnetic field near the magnetic neutral line would be missed in the measurement, where stronger shearing motions are generally found.

Due to the limitations of the vector magnetograms in last decade, time series of line-of-sight magnetograms have been used for statistical studies on magnetic helicity injection of active regions (e.g. LaBonte et al. 2007; Jeong & Chae 2007; Tian & Alexander 2008 and 2009; see a review paper of Démoulin & Pariat 2009, and references therein). The MDI/96m full disk data have been routinely obtained by the Michelson Doppler Imager (MDI) on the Solar and Heliosphere Observatory (SOHO) since 1996, with a large time interval (96 minutes) and a low spatial resolution (2 arcseconds). MDI line-of-sight magnetograms with high spatial resolution (0.6 arcseconds) and short time interval (1 minute) have also been obtained for specific time intervals.

The local tracking methods are designed to follow the photospheric footpoints of flux tubes based on the line-of-sight magnetograms, e.g. LCT (see Chae & Jeong 2005), Fourier LCT (FLCT, Welsch et al. 2004), and differential affine velocity estimator (DAVE, Schuck 2005, 2006). The footpoint velocity has two contributions, a first one due to the true hor-izontal plasma flow velocity ($\mathbf{v}_h$) and a second one due to the vertical motion of the flux tube ($v_n$). By a simple geometric argument, the vertical motion induces an apparent motion with a velocity equal to -($\frac{v_n}{B_n}$)$\mathbf{B}_h$. Then the footpoint velocity, $\mathbf{u}$, is the sum of these two horizontal velocities (see Démoulin & Berger, 2003),

$$\mathbf{u} = \mathbf{v}_h - \frac{v_n}{B_n}\mathbf{B}_h. \tag{2}$$

Therefore, Equation (1) can be rewritten as,

$$\frac{dH}{dt} = -2\int_S (\mathbf{A}_p \cdot \mathbf{u})B_n \, \mathrm{d}S = \int_S G_A \, \mathrm{d}S. \tag{3}$$

Here $G_A = -2(\mathbf{A}_p \cdot \mathbf{u})B_n$ is defined as the magnetic helicity flux density. With the approximation of a photosphere locally planar, an explicit expression of $A_p$ is (Berger, 1984)

$$A_p(\mathbf{x}) = \frac{1}{2\pi}\hat{\mathbf{n}} \times \int_{S'} B_n(\mathbf{x}')\frac{\mathbf{r}}{r^2} \, \mathrm{d}S',$$

where $\mathbf{r} = \mathbf{x} - \mathbf{x}'$ is the vector between the two photospheric positions defined by $\mathbf{x}$ and $\mathbf{x}'$.

However, Pariat et al. (2005, 2006) found that $G_A$ has 'false' bipolar polarities even in the simple case of a magnetic flux tube moving with a spatially uniform velocity. In order to solve the problem, Equation (3) was transformed to,

$$\frac{dH}{dt} = -\frac{1}{2\pi}\int_S\int_{S'} \frac{d\theta(\mathbf{r})}{dt} B_n B_n' \, \mathrm{d}S \, \mathrm{d}S' = \int_S G_\theta(\mathbf{x}) \, \mathrm{d}S, \tag{4}$$

where $d\theta/dt$ is defined by

$$\frac{d\theta(\mathbf{r})}{dt} = \frac{1}{r^2}\left(\mathbf{r} \times \frac{d\mathbf{r}}{dt}\right)_n = \frac{1}{r^2}(\mathbf{r} \times (\mathbf{u} - \mathbf{u}'))_n \, ,$$

and $G_\theta$ is defined by

$$G_\theta(\mathbf{x}) \;=\; -\frac{B_n}{2\pi}\int_{S'} \frac{d\theta(\mathbf{r})}{dt} B_n' \, \mathrm{d}S'. \tag{5}$$

More details of the derivation can be found in the paper of Pariat et al. (2005). Since its expression depends on the relative velocity between two points, the 'false' signal appearing in the map of $G_A$ is removed. More generally, it was shown that $G_\theta$ is a better proxy of the magnetic helicity flux density (see more details in Pariat et al. 2005, 2006; Chae 2007).

In this paper, we use the DAVE technique (Schuck 2006) on time series of MDI/1m line-of-sight magnetograms with high spatial resolution to measure the velocity field $\mathbf{u}$, and then to calculate magnetic helicity injection rate based on Equations (4) and (5). One of the motivations is to determine if the magnetic helicity between the leading and following polarities is asymmetric, as found by Tian & Alexander (2009). They used an LCT technique on time series of MDI/96m line-of-sight magnetograms with lower spatial resolution to calculate magnetic helicity injection rate and flux based on Equations (2) and (3). Recently, the simulations of Fan et al. (2009) theoretically support the presence of the magnetic helicity asymmetry, which showed a number of features consistent with the observational results of Tian & Alexander (2009). Fan et al. (2009) found that the field lines in the leading leg show more coherent values of the local twist, whereas the values in the following leg show larger fluctuations with mixed signs. Furthermore, a faster rotation of the leading polarity sunspot is found, contributing to a greater helicity injection rate in the leading polarity of an emerging active region. This result is coherent with the simplified model of active region flux emergence into the corona proposed by Longcope & Welsch (2000) when one takes into account the difference of magnetic field strength between the leading and following polarities.

We organize our paper with the following structure: Section 2 introduces data and the DAVE technique. Sections 3, 4 and 5 describe the active region sample used in the paper (three newly emerging active regions), and show results associated with them. Discussions and conclusions will be given in Sections 6 and 7.

## 2. Data and Method

In this paper, three newly emerging active regions (ARs: NOAA 10365, 10319, and 10323) have been investigated to study the properties of magnetic helicity injection. The ARs were selected because they have a large data set of lev1.5 MDI/1m line-of-sight magnetograms with a spatial resolution of 0.6 arcseconds. In order to compare with previously published results (e.g. Tian & Alexander 2009), we also use lev1.8.2 MDI/96m line-of-sight magnetograms with a spatial resolution of 2.0 arcseconds. The three ARs were isolated and had a well-developed predominantly bipolar magnetic structure. Though AR 10365 shows some magnetic complexity, however, it is still predominantly bipolar. Therefore, all positive/negative flux ($|B| \geq 20G$) belongs approximately to the leading/following polarity for AR 10319 (N13) in the northern hemisphere, while the opposite applies for ARs 10365 (S08) and 10323 (S09) in the southern hemisphere. We notice that AR 10365 has the peculiarity to emerge in a north-south direction and later on to have the leading magnetic polarity moving eastward of the following polarity (Chae et al. 2004, also see Section 3.1). The emergence and development of the three active regions will be described in detail in the following section.

The MDI line-of-sight magnetograms were first corrected into a radial direction, based on a method introduced by Chae & Jeong (2005), which was used in previous papers of Tian & Alexander (2008, 2009). The magnetic field was also corrected by the standard multiplier of 1.6, to account for an underestimation found in original lev1.5 MDI data (see

Berger & Lites 2003). Note that lev1.8.2 MDI/96m data do not need this enlarging. The local coordinates in the image plane have been transferred to heliographic coordinates, and the flux-weighted centers of the two opposite polarities and their polarity separation (d) are calculated. The total positive and negative radial magnetic flux ($\Phi$) across an active region is calculated where the magnetic field $|B_n| \geqslant 20G$.

Differential affine velocity estimator (DAVE) is designed and developed to measure photospheric mass flow velocity (Schuck 2005, 2006). The technique involves a short time-expansion of the modified LCT method for estimating the velocity field from the time series of line-of-sight magnetograms. The method adopts an affine velocity profile describing a velocity field inside a local area around a specified point. It has been shown to produce the velocity field much better than the LCT method in regions, like vortexes and saddle points, where the spatial variation of the velocity is important (see Schuck 2006; Chae 2007). The method requires that the velocity field ($\mathbf{u}$) should satisfy the induction equation for $B_n$ in each local window,

$$\partial_t B_n + \nabla_h \cdot (B_n \mathbf{u}) = 0. \tag{6}$$

Here, $\partial_t B_n$ is the local evolution of vertical component of the magnetic field ($B_n$); 'h' and 'n' denote horizontal and vertical (normal) directions. This guarantees that the determined velocity $\mathbf{u}$ has a clear physical meaning, representing the apparent motion of fieldline foot-points in the photosphere. Schuck (2006) demonstrated that this technique was faster and more accurate than existing LCT algorithms (see also Chae 2007).

The DAVE method is a local optical flow method that determines the mass velocities within a windowed subregion by constraining the local velocity profile. Therefore, the choice of window size is crucial to determine the velocities accurately. The window size must be large enough to contain enough structure to uniquely determine the coefficients of the flow profile and to resolve the aperture problem, but not so large as to violate the affine velocity profile which is only valid locally (see Schuck 2006, 2008). Schuck (2008) found that a square window of approximately 20 pixels provided the best performance. Since the region of the three ARs is rectangular (see Figures 1a, 6a, and 10a), we prefer to follow Welsch et al. (2007), and choose rectangular windows. We select four window sizes, $6 \times 5$, $12 \times 10$, $18 \times 15$, and $24 \times 20$ pixels, in order to test which one performs best when using the MDI/1m high spatial resolution line-of-sight magnetograms. In using the DAVE technique, another free parameter is the time interval of the two images tracked. It has an important effect in determining the photospheric flow motion velocity (Schuck 2006, 2008). As a consequence, the determination of the magnetic helicity injection rate has been tested with four time intervals $\Delta t$=5, 10, 30, and 60 mins for the MDI/1m magnetograms.

In this paper, we employ $G_\theta$ defined in Equation (5) to be the proxy of magnetic helicity flux density. This choice is motivated by the fact that $G_A$ (defined by Equation (3)) introduces 'false' signal, especially for the leading polarity with big and compact sunspots (see Pariat et al. 2006; Nindos et al. 2003; Tian & Alexander 2008). $G_\theta$ was introduced to remove those 'false' bipolar polarities. $G_\theta$ is calculated by using a convolution technique based on the fast Fourier transform (FFT), introduced by Chae (2007). $G_\theta$ is then summed up separately for the positive and negative magnetic polarities to provide the magnetic helicity flux in both magnetic polarities. We find that some anomalously large velocities appear in very weak field regions far from the main active region when the DAVE tecjnique is used (see also Figure 7 of Jeong & Chae 2007; Figure 1 of Chae 2007). We assume that those large velocities are not 'real', and calculate the magnetic helicity injection rate when the apparent speed $u \leq 0.5$ km/s in this paper ($u$ is defined in Equation (2), see also Figure 3 of Welsch et al. 2009). We found that the contribution from magnetic structures with speeds larger than 0.5 km/s is negligible, thus the limit on $u$ ($\leq 0.5$ km/s) has no significant effect on our results.

## 3. Emerging Active Region: NOAA 10365 (S08)

### 3.1. Active Region Evolution

Full disk MDI/96m line-of-sight magnetograms show that the emergence of a new bipolar region with strong field occurred at about S08E40 between 2003 May 24 and 25, named NOAA 10365. The active region had a roughly north-south orientation with negative polarity in the north and positive polarity in the south producing a large tilt towards the equator (about 90°) at the beginning of its emergence (see Figure 1 in Chae et al. 2004). With the flux emergence, the active region continuously rotated counter-clockwise. Based on the Hale and Nicholson polarity law and Joy's Law (Hale et al. 1919; Zirin 1988), most bipolar active regions have a small tilt angle (in general $< 40°$, see also Tian et al. 2005, and references therein), and its leading polarity has negative flux, while the following polarity has positive flux for the active regions located at the southern hemisphere in solar cycle 23. Therefore, NOAA 10365 is peculiar since it had a large tilt to the equator, and moreover it rotated away from the east-west orientation present for the majority of active regions (Figure 1a). A minority of active regions have large tilt and/or such kind of rotation (see López Fuentes et al. 2003; Tian et al. 2005).

From MDI/1m line-of-sight magnetograms over 3 days shown in Figure 1a, we see that the magnetic field pattern of the emerging bipole is characterized by two elongated areas of opposite polarity, called 'magnetic tongues' by López Fuentes et al. (2000, see also Chandra et al. 2009 for this active region). The tongues are present only when the apex of the twisted flux tube is crossing the photosphere during the flux emergence, because of the contribution of the azimuthal component of the emerging twisted flux tube to the observed vertical component of the photospheric field. According to the geometric shape of the magnetic tongues, the active region 10365 should have a twisted flux tube with positive helicity in agreement with the magnetic helicity injected at the photospheric level (see Figure 2; Chae et al. 2004; Jeong & Chae 2007; LaBonte et al. 2007). With the flux emergence, the positive and negative magnetic flux and the separation distance of the two polarities roughly increased linearly from the start of the emergence up to the time '$t_d$' (see Figure 1b and 1c), which denotes the major period of strong emergence of the flux tube. Later, the polarity separation decreased sharply.

In Figure 1b, the bold ($P_1$) and dotted ($N_1$) curves display 1.6 times the positive and negative magnetic flux calculated by MDI/1m magnetograms. After the time '$t_d$', the measured flux is about $5 \times 10^{21}$ Mx larger than that calculated from the MDI/96m magnetograms (2.0 arcsecs per pixel), denoted by '$\diamond$' (and $P_0$ for positive flux) and '$*$' (and $N_0$ for negative flux). The flux difference is already evident after time '$t_0$', when the active region was crossing the central meridian line. This comparison reflects a large underestimation of the flux measured from the MDI/96m data in the strong field region due to the much poorer spatial resolution.

### 3.2. Magnetic Helicity Injection and Asymmetry of Leading and Following Polarities

Using the same threshold (20 G) and approximately the same apodizing window size (in arcsecs), we find in Figure 2 that the magnetic helicity injection rate (dH/dt) and magnetic helicity flux ($H$) calculated by MDI/96m magnetograms is significantly lower (about 2 times) than that calculated by MDI/1m magnetograms (see panels a1 and b1, a3 and b3). However, the 'normalized' helicity rate ($\frac{dH/dt_P}{\Phi_P^2}$ and $\frac{dH/dt_N}{\Phi_N^2}$) shown in Figure a2 and b2 has no significant difference. Here, the time interval ($\Delta$t) of the two images used is 96 minutes and 10 minutes, respectively: these quantities are used in cross-correlation analysis to calculate the photospheric flow velocity in the DAVE technique (see Section 2).

The magnetic helicity injection rate of the two polarities has similar temporal profiles, but shifted in magnitude ('$\diamond$' for the positive polarity and '$*$' for the negative polarity). This is present both with MDI/1m data and 96m data (comparing the first and second rows in Figure 2). Between the times '$t_1$' and '$t_d$', the 'normalized' helicity rate has roughly the same linear-trend with a slope of $K = 0.1$ (in units of $4{\times}10^{-4}\text{h}^{-2}$) for the two polarities, denoted by the thick lines. Prior to time '$t_1$', however, there is a large difference at the beginning of the flux emergence. This is possibly due to large uncertainties in the measurements during the early emerging phase. After '$t_d$', the injection rate significantly decreased, and this is well related to the sudden decrease of the polarity separation (see Figure 1c). Since the major flux emergence occurred before '$t_d$' (Figure 1), it is clear that the magnetic helicity injection is strongly associated with the flux emergence of the active region.

The significant asymmetry of the injection rate between the positive and the negative polarities before '$t_d$' implies an important asymmetry of the magnetic helicity injected (panels a3 and b3 in Figure 2). The asymmetry is about a factor 2 by the end of the analyzed period, when the injected magnetic helicity almost saturates to a maximum value. This asymmetry is further demonstrated in Figure 3 where the x-axis is the magnetic helicity injection rate and the normalized helicity injection rate of the negative (leading) polarity, $dH/dt_N$ and $dH/dt_N/\Phi_N^2$; and the y-axis shows the equivalent properties for the positive (following) polarity, $dH/dt_P$ and $dH/dt_P/\Phi_P^2$. The slope values are $0.5-0.6$ with reasonably good correlation coefficients except for the 'normalized' magnetic helicity with MDI/96m (see panel a2). Though the amount of the injection rate shown in panel a1 is about two times smaller than that shown in panel b1, the asymmetry (see slope values 'K') are very similar with MDI/96m and MDI/1m data for the helicity injection rate. Taking into account the possible effect from a measured magnetic flux imbalance, we remove this effect by using the 'normalized' helicity injection rate ($dH/dt_N/\Phi_N^2$ and $dH/dt_P/\Phi_P^2$, see panels a2 and b2). However, we found no tendency for positive (following) polarity because of the low correlation coefficient (see $C_r$ in panel a2).

### 3.3. Dependence of Magnetic Helicity Injection on Apodizing Window Size and Time Interval

In most of calculations of magnetic helicity injection rate using the MDI/1m data, we generally choose a window size of $18 \times 15$ pixels, which was estimated to be around the best choice to obtain the flow velocity within the constraints of the DAVE technique (see detailed description in Schuck 2006, 2008). We also choose three other window sizes for comparison: $6 \times 5$, $12 \times 10$, and $24 \times 20$ pixels. From Figure 4, we see that the result at a window size of $12 \times 10$ pixels is very close to that at the window size of $18 \times 15$ pixels (see the correlation coefficient '$C_r$' and the slope 'K' of the linear-fit line in the panel a2). The scatter is the biggest at the window size of $6 \times 5$ pixels (see the low '$C_r$' in the panel a1), while the rate is 6% lower at $24 \times 20$ pixels (see the slope 'K' in the panel a3). Generally speaking, the magnetic helicity injection rate depends little on the window sizes, and the optical window size is expected to be between $12 \times 10$ and $18 \times 15$ pixels.

In using the DAVE technique, another important parameter is the time interval of two images tracked. It is noted that in most of our calculations using the MDI/1m data, the time interval $\Delta t$ chosen is generally 10 min. However, in Figure 5, we show the magnetic helicity injection rate at $\Delta t$=5, 30 and 60 min, and compare them with that at $\Delta t$=10 min, as well as comparing them with MDI/96m data at $\Delta t$=96m.

Because the data points are different between the compared magnetic helicity injection rate at $\Delta t$=10 min and $\Delta t$=5, 30, 60, and 96 min within the same period, an interpolation is needed. The simplest approach is to use a linear interpolation. A more sophisticated technique is to use a cubic spline interpolation, which employs a cubic polynomial between each pair of consecutive data points with the constraint that the polynomial crosses both data points, and that there is the same first and second derivative on both sides (of each data point). So a cubic spline interpolation produces a smooth interpolating curve, then we can interpolate the quantity along the y abscissa and compute it at data points of the x abscissa. In our case, since there are many points, relatively well ordered, the above plots are very similar for the linear and spline interpolations, so the interpolation method is not an issue.

Such plots based on the cubic spline interpolation are shown in Figure 5 with the results for the positive polarity in panels (a1−a4) and those for the negative polarity in panels (b1−b4). We find that the slope values ($K$) of the linear-fit lines decrease with the increase of the time interval, and the correlation coefficients ($Cr$) also generally decrease. The rates at $\Delta t$=5m have larger scatters, and the correlation coefficient is worse than that at $\Delta t$=30m. The results described above reflect a large dependence of the magnetic helicity injection rate on the selection of the time interval of the two images tracked. From Figure 5 (panels a3−b3), it is found that the helicity injection rate is a factor $\approx$ 0.69 lower for $\Delta t$=60m than for $\Delta t$=10m. The case with $\Delta t$=60m does not have the injection rate at the lower time cadences, so it misses the power at all time scales lower than 60m (particularly at 5m and 10m). The case $\Delta t$=5m has $K \approx 1$ (see panels a1−b1), so there is no more flux than at $\Delta t$=10m. Therefore, these results imply that magnetic structures with a short lifetimes, 10 − 60m, are contributing about 30% to the coronal helicity injection.

## 4. Emerging Active Region: NOAA 10319 (N13)

### 4.1. Active Region Evolution

Based on activity report at Big Bear Solar Observatory (BBSO), NOAA 10319 appeared at the east limb on 2003 March 22, with a small size. It was an emerging active region and grew to be a simple $\beta$ region from March 24. The region developed rapidly with an increase in sunspot area on March 27. From March 29, the large $\beta$ region was in a decaying stage. MDI/1m line-of-sight magnetograms over three days close to the central meridian line are shown in Figure 6a, where the flux emergence and its growing evolution are evident. The active region is a normal region in the northern hemisphere, with positive/negative polarity as the leading/following (Hale et al. 1919; Zirin 1988). Moreover, it had an roughly east−west orientation with a bipolar titled about 15−25° to the equator and with the leading magnetic polarity closer to the equator (see Figure 6a).

Figure 6b and 6c clearly display the emergence proceeding over the three days. Based on MDI/1m data, the magnetic flux linearly increases from March 27, shown as bold (dotted) curves (see Figure 6b). However, the flux calculated by MDI/96m data, shown as '$\diamond$' and '$*$', does not capture this fast increase, and rather shows an artificial saturation. This again reflects that the MDI/96m data significantly underestimate the magnetic flux in strong field regions, as found in NOAA 10365 (see Figure 1b). The polarity separation (see Figure 6c) increases linearly, so this case is compatible with the simple emergence of an $\Omega$ flux tube.

The most notable properties, here, are the balanced magnetic flux between the leading (positive) and following (negative) polarities, but asymmetric magnetic morphology: compact leading polarity while dispersed and fragmented following polarity. The magnetic field pattern of the emerging bipolar structure is also characterized by two elongated areas of opposite polarity, magnetic 'tongues' (see López Fuentes et al. 2000). The geometric shape of the magnetic tongues on March 26 indicates positive helicity in the emerging twisted flux tube (see more details in Section 3.1).

### 4.2. Magnetic Helicity Injection

We performed a similar calculation and analysis to those shown in Section 3.2 and Figure 2. The results are shown in Figure 7, where we find a similar temporal trend between the MDI 96m (see panels a1−a3) and 1m data (see panels b1−b3). However, the amount is significantly different with at least 2 times smaller magnetic helicity injection rate for 96m data. The magnetic helicity injection rate is very asymmetric between the two polarities (see '$\diamond$' and '$*$'), though their magnetic flux is almost balanced (see Figure 6b). Before time 't1', both magnetic helicity injection rates are positive with MDI/1m, but with the leading (positive) polarity being about 2−3 times larger. However, at later times, the leading polarity injects negative magnetic helicity.

The results of similar calculations and analysis to those shown in Section 3.3 and Figures 4 and 5 are shown in Figures 8 and 9, respectively. Essentially we reach the same conclusions and summarized as:

(1) The results at the window sizes of $12 \times 10$ and $24 \times 20$ pixels are the closest to that at $18 \times 15$ pixels, the value identified by Schuck (2006) as being the optimal window size, while the scatter and deviation are the largest at $6 \times 5$ pixels (Figure 8).

(2) The slopes ($K$), of a linear fit between results obtained with different $\Delta t$, decrease with the increase of the time interval ($\Delta t$), and the correlation coefficients ($Cr$) also generally decrease for both magnetic polarities (Figure 9). The largest scatter and a low correlation coefficient appear at $\Delta t$=5m.

(3) The slopes and the correlation coefficients of positive (leading) polarity are generally better than that of negative (following) polarity (Figure 9).

These again reflect little dependence of the helicity injection rate on the window size selected, but with a large dependence on the time interval selected. From these results we conclude that the window size between $12 \times 10$ and $18 \times 15$ pixels, and the time interval of $\Delta t$=10m are probably the best ones to compute the magnetic helicity injection from MDI/1m magnetograms. As in Section 3.3, we conclude the magnetic structures with short lifetimes, $10 - 60$m, contribute significantly to the magnetic helicity injection (30−50%).

## 5. Emerging Active Region: NOAA 10323 (S09)

### 5.1. Active Region Evolution

NOAA 10323 was a very small active region emerging on the eastern disk on 2003 March 25, and grew slowly over the following three days. On March 29, the region has grown rapidly, becoming an $\beta$–$\gamma$ region. It slightly decayed in sunspot area from April 1. The MDI/1m line-of-sight magnetograms over three days close to central meridian line are shown in Figure 10a, where the flux emergence and its growing evolution are most evident.

From Figure 10b, it is found that the magnetic flux linearly increases, where bold and dotted curves denote results calculated from MDI/1m data, and '$\diamond$' and '$*$' denote results calculated from MDI/96m data. The amount of the magnetic flux found from the MDI/1m is much smaller than the one found from MDI/96m before March 30. This behavior was not seen in the two other active regions shown in Figures 1b and 6b. It is noted here that the magnetic flux of the two polarities is approximately balanced on the first day, while displaying larger flux in the leading (negative) polarity after that. As most of bipolar active regions, the magnetic morphology is asymmetric with more compact magnetic field in the leading side while more dispersed and fragmented field in the following side. The polar separation increased first, and then suddenly decreased over abou 2 days (see dashed−dotted lines in Figure 10c).

Magnetic 'tongues' are also present in this active region (the positive magnetic polarity is present northward to the negative polarity in the center of the active region). This indicates the emergence of a twisted flux tube with positive magnetic helicity, in agreement with the helicity flux measured in both polarities with MDI/1m (e.g. panel b3 of Figure 11).

### 5.2. Magnetic Helicity Injection

The magnetic helicity injection rate is shown in Figure 11, where the negative (leading) polarity (shown by $*$) injects positive helicity with typically larger values than the positive (following) polarity (shown by $\diamond$) which injects helicity with a mixed sign over most of the time considered here. The helicity injection rate is very small before March 30, but large after that, especially for the leading (negative) polarity. The MDI/96m data display two times smaller injection of magnetic helicity than the MDI/1m data for the leading polarity, and showing an opposite sign in the following polarity (see panels a3 and b3 in Figure 11). Therefore, we again note that the MDI/96m data are not suitable to be used for quantitative tracking studies or for determining the amount of the magnetic helicity injected.

When we study the dependence of the magnetic injection rate of the active region on the window size and the time interval, shown in Figures 12 and 13, the same conclusions are obtained to that shown in Figures 4 and 8, 5 and 9, and Sections 3.3 and 4.2 (see more details there).

## 6. Discussion

We summarize in Table 1 and Figure 14 the slope (K) of the linear least square fit between results with various $\Delta t$ values compare to those at $\Delta t$ = 10m. Such fits are shown in Figures 5, 9, and 13. From Table 1, for $\Delta t \geq$ 30m, we found that the values of the slope ($K$), and thus the magnetic helicity injection rate, decreases with the increase of the time interval $\Delta t$ selected for the three active regions. The injection rates at the smallest time interval, $\Delta t$=5m, are generally lower than that at $\Delta t$=10m due to $K < 1.0$, while the rates at $\Delta t \geqslant$ 30m are much worse, denoted by much lower $K$. Therefore, the largest value of the injection rate should be obtained when the time interval is close to 10m. This is also reflected in Figure 14a, which plots the slopes ($K$) of the linear-fits versus the correlation coefficients ($Cr$) at $\Delta t$=5, 30, 60, and 96m for the two polarities of the three active regions. It is found that the slopes are closer to 1.0, but the correlation coefficients are generally bad at $\Delta t$=5m (see '+') since there is a large dispersion in the measurements (Figures 5, 9, and 13). The slope values are far from 1.0 at $\Delta t$=60m (see '□'). Of course, the worst results are obtained by using the MDI/96m data at $\Delta t$=96m, with both low slopes and bad correlation coefficients (see '×').

The performance of each active region is shown in Figure 14 (b1−b3), where $\diamond/*$ denotes the values related to the positive/negative magnetic flux. In general, the slope close to 1.0 and having a better correlation coefficient is at the time interval close to 10m. For each active region, shown in Figure 14b, the leading polarities generally have better slope values and correlation coefficients than the following polarities (see also the values 'K' of leading (L) and following (F) polarity in Table 1).

### 6.1. Dependence on the Window Size and the Time Interval

The strong dependence on the time interval is expected, since the DAVE technique is the most accurate when the magnetic field moves on the order of a pixel between frames. For larger displacement, reliable results are not obtained. This is the Courant Condition also called the Courant-Friedrichs-Lewy condition (derived in Schuck 2006 for a simple case). However, Chae et al. (2004) did not find a dependence of the magnetic helicity injection rate on the time interval. They used a different technique, called LCT, on MDI/1m line-of-sight magnetograms for AR 10365, and tested the injection rate at the time interval $\Delta t$= 5m and 15m. Their results indicated that the magnetic helicity injection rate is comparable between using MDI/1m and 96m data, (see Figure 8 of Chae et al. 2004). However, their comparison of MDI/1m and 96m data is limited to a time period of $\approx$ 5 hours on May $27^{th}$, while our comparison is realized over $\approx$ 2.5 days (see Figures 2 and 5).

Table 1: Summaries of slopes ($K$) shown in Figures 5, 9, and 13. The window size is 18×15 pixels. L/F denotes leading/following polarity, while P/N denotes positive/negative magnetic flux. 1 $\sigma$ standard error is for the slope of the least-square fit.

| NOAA | Each Polarity | Slopes ($K\pm$ 1$\sigma$) | | | |
|---|---|---|---|---|---|
| | | $\Delta t$=5m | $\Delta t$=30m | $\Delta t$=60m | $\Delta t$=96m |
| 10365 (S08) | L/N | 0.97±0.02 | 0.87±0.01 | 0.69±0.02 | 0.39±0.01 |
| | F/P | 1.00±0.02 | 0.81±0.02 | 0.69±0.02 | 0.52±0.02 |
| 10319 (N13) | L/P | 0.88±0.08 | 0.83±0.02 | 0.69±0.02 | 0.36±0.01 |
| | F/N | 0.89±0.07 | 0.69±0.03 | 0.50±0.02 | 0.10±0.01 |
| 10323 (S09) | L/N | 1.00±0.03 | 0.81±0.02 | 0.54±0.02 | 0.64±0.02 |
| | F/P | 0.96±0.03 | 0.68±0.03 | 0.53±0.03 | 0.05±0.05 |

We found that a 96m interval is too long for DAVE, so that this is possibly one reason for the 96m data to have the smallest slopes and correlation coefficients shown in Figures 5, 9 and 13. Other reasons are the low spatial resolution and large underestimation of flux in strong field regions. The results are much worse for the following polarities of the two last ARs, since they have more dispersed fields than the leading polarities. These imply that the 96m data are not suitable for detailed tracking studies and for determining quantitatively the magnetic helicity injection. However, it should be noted, that the 96m data can be useful for showing the trends, e.g. helicity asymmetry and time development.

It is not surprising that the dependence of magnetic helicity injection rate on the window size is weak. The DAVE technique allows the velocity profile to vary somewhat within the aperture, thereby, mitigating the window size dependence. Since the DAVE technique uses finite differences to obtain the spatial and temporal derivatives, it works best when the displacement between the two images tracked is less than 1 pixel, and it might be worse with larger displacements (see more details in Schuck 2006, 2008; Chae 2007; Chae & Sakurai 2008). Therefore, a small window size and a small time interval would be the optimal choice for the tracking study. However, the smallest window size tested (6×5), and the shortest time interval tested (5m), would produce large scatters. Our tests in this paper demonstrate that the window size within a region of 12×10 and 18×15 pixels, and a time interval close to 10m would be the best for the calculation of magnetic helicity injection rate using DAVE and MDI/1m data.

### 6.2. Effect of the Noise on the Results

Diverging flows are often observed near the neutral line of opposite magnetic polarities during the flux emergence of an active region. Shearing velocity vectors point along the normal-field contours, and converging or diverging flows near the neutral line are parallel or anti-parallel to the normal-field gradients there. These motions are stronger during the flux emergence, and some of the motions may be missed in the calculation of magnetic helicity injection, when the threshold of the magnetic field is chosen to be $20G$, as the measurements shown in this paper. We estimate the possible effect resulting from the threshold selected by comparing the magnetic helicity rate calculated for magnetic field including and excluding magnetic field lower than $20G$. Figure 15 shows the deviation of the magnetic flux and the magnetic helicity rate of the magnetic field over 20 G and 0 G. From the Figure, it is found that the deviation of the magnetic flux is about 3−5%, and the difference of the magnetic helicity injection rate is generally lower than 10%. From this we can infer that the effect of the threshold selection and some probable contributions from the shearing and diverging motions are negligible in low field regions.

### 6.3. Line-of-sight tracking techniques

In this paper, we used the line-of-sight tracking technique, DAVE, to measure photospheric flow velocities, and then to calculate the magnetic helicity injection rate. In the presence of vertical flows and horizontal magnetic fields, Schuck (2008) argued that the line-of-sight tracking methods do not accurately capture the complete footpoint dynamics. However, Schuck (2008) used an anelastic pseudo-spectral ANMHD simulation for that test, which models the rise of a buoyant magnetic flux rope in the convection zone and represents the structure of granulation or super-granulation, rather than the dynamics of an active region. On the other hand, Welsch et al. (2007) noted that the tracking methods performed better on real magnetograms than on the synthetic ANMHD data. We agree that not all magnetic helicity flux can be measured using the line-of-sight tracking methods, limiting the results of such analysis. In particular, Démoulin & Berger (2003) argued that motions along the iso-contour of $B_n$ (like twisting motions) could not be detected, by any method with longitudinal magnetograms. However, following the results of Démoulin & Berger, in particular the derivation of Equation (3), we claim that the tracking methods can capture the horizontal plasma motions, and at least some part of the projected transverse velocity resulting from the emergence of an active region as an $\Omega$-shaped flux tube. The measurement of the horizontal velocity and magnetic helicity injection rate based on it must carry part of the message and characteristics of the magnetic emergence, as evidenced by the studies of Jeong & Chae (2007) and Tian & Alexander (2008), who found that there is very close relationship of the magnetic helicity injection rate (dH/dt) and the flux emergence, and the results were also found to agree with the results of models with flux emergence (see more details in Section 3.2 of Démoulin & Pariat 2009; see also Fan et al. 2009).

## 7. Conclusions

In this paper, we investigate three newly emerging active regions (NOAA 10365, 10319, and 10323), having the most data of MDI/1m line-of-sight magnetograms with high spatial resolution. The main aim is to check if the magnetic helicity injected is asymmetric between the leading and following polarities as it was found by Tian & Alexander (2009) with MDI/96m data. In order to calculate the magnetic helicity injection rate of each polarity, we employ a well-developed technique (DAVE) for tracking photospheric flow motion. Comparable results are found from the three active regions:

(1) An important magnetic helicity asymmetry is present between the leading and following polarities. The asymmetry is present in the magnetic helicity injection rate ($\mathrm{d}H/\mathrm{d}t$), the 'normalized' helicity injection rate ($\mathrm{d}H/\mathrm{d}t/\Phi^2$), and the magnetic helicity flux ($H$) of each polarity (see Figures 2, 7, and 11).

(2) The magnetic helicity injection rate ($\mathrm{d}H/\mathrm{d}t$) depends little on the window size selected to perform the cross-correlation (see Figures 4, 8, and 12). The best performance is at a window size of 12×10 and 18×15 pixels. The magnetic helicity injection rate has a large dispersion when the window size is small, e.g. 6×5 pixels.

(3) The magnetic helicity injection rate ($\mathrm{d}H/\mathrm{d}t$) depends more on the time interval $\Delta t$ of two images tracked (see Figures 5, 9, and 13). In general, the shorter the time interval is, the larger the injection rate is. With the presence of increasing dispersion for low $\Delta t$, we conclude that the most appropriate time interval is close to 10 minutes. It is found that magnetic structures with a short-life, $10-60$m, could contribute about $30-50\%$ to magnetic helicity injection rate (see Figures 5, 9, and 13).

(4) The comparison with the results calculated by MDI/96m data with a spatial resolution of 2 arcseconds shows that the MDI/96m data can reflect the general trend of the magnetic properties, especially for the leading polarity. However, the magnetic flux is significantly underestimated in strong field region, and the 96m data are not suitable for quantitative tracking studies, so to precisely estimate magnetic helicity injected.

From a statistical study of 13 emerging active regions, Tian & Alexander (2009) found that the magnetic helicity is asymmetric between the leading and following polarities. In their calculation of magnetic helicity injection rate, they employed an LCT technique (see Chae & Jeong 2005), which determines an average velocity over the apodizing window selected to perform the cross-correlation (see Welsch et al. 2007; Chae & Sakurai 2008 for more discussions). Thus, there could be a significant relative error when the velocity field has important spatial gradients. On the other hand, they used MDI/96m line-of-sight magnetograms, for which the spatial resolution is low (2 arcseconds). Moreover, the magnetic flux is significantly underestimated in the strong field region and the time interval (96m) is too long for realistic tracking studies, based on the results reported here. However, the data may be used for characteristic study, because we find that the 96m data can follow the trend of magnetic property of magnetic helicity injection (see Figures 2, 7, and 11). Moreover, we have confirmed the existence of magnetic helicity asymmetry between the leading and following polarities, using a well-developed tracking technique (DAVE) on a time series of MDI/1m data with high spatial resolution.

Fan et al. (2009) have examined the twist distribution as well as magnetic helicity flux from the rise of an asymmetric Ω-shaped tube, based on the anelastic MHD simulation of Fan (2008). They found that a natural consequence of the significantly stronger field strength in the leading leg of the Ω-tube compared to the following leg, is that the field lines in the leading side wind about each other more smoothly, showing more coherent values of the local twist, whereas the field lines in the following leg are more frayed and show significantly larger fluctuations and mixed sign of the local twist. They argued that the faster rising of the leading leg than the following leg results in a greater vertical helicity flux in the leading polarity. This may contribute to the asymmetry of the magnetic helicity injection observed in emerging active regions (see also Tian & Alexander 2009). Another important effect could result from the greater Alfvén speed along the leading tube resulting from its stronger field strength, which would drive a faster rotation of the leading spot and hence produce a greater helicity injection in the leading polarity. This simulation showed a number of features consistent with the observational results reported in this paper.

An important question related to the observations presented in this paper is whether the derived asymmetries in the magnetic helicity are reflected in some way in the corona. The difficult here lies in separating the consequences of the asymmetries in the flux distributions from those in the helicity distributions. Given the high coronal Alfven speed, it is expected that any helicity asymmetry will be rapidly redistributed within the corona ( minutes). For the amount of energy release and associated localization of the reconnection regions, the complexity of the magnetic topology is expected to be more important than the asymmetry of the helicity injection. Therefore, the observations and analysis of the asymmetry of the helicity injection more generally provide constraints on the emergence process, and on the physics involved at and below the photosphere, rather than directly leading to large-scale coronal phenomena. However, the complex distribution of thin flux tubes with mixed helicity in the following polarity may lead to local magnetic configurations conducive to the formation of many small-scale current sheets where magnetic reconnection can drive the production of small-scale transients, e.g. bright points and small flares. The coronal consequences of the asymmetric helicity injection and flux emergence will be investigated in the future.

The authors are grateful to MDI teams for the wonderful data. L. Tian thanks Drs. Leka, Pevtsov, and Schuck for their comments and suggestions during this study. This research is supported by NASA grant NNX07AH38G.

## REFERENCES

Berger, M. A., & Field, G. B. 1984, J. Fluid, Mech., 147, 133

Berger, T., & Lites, B. 2003, Sol. Phys., 213, 213

Chae, J., Moon, Y., & Park, Y. 2004, Sol. Phys., 223, 39

Chae, J., & Jeong, H. 2005, Journal of Korean Astronomical Society, 38, 295

Chae, J. 2007, Adv. Space Res., 39, 1700

Chae, J., & Sakurai, T. 2008, ApJ, 689, 593

Chandra, R., Schmieder, B., Aulanier, G., & Malherbe, J. M. 2009, Sol. Phys., 258, 53

Démoulin, P., & Berger, M. 2003, Sol. Phys., 215, 203

Démoulin, P. 2007, Adv. Space Res., 39, 1674

Démoulin, P. & Pariat A. 2009, Adv. Space Res., 43, 1013

Fan, Y. 2008, ApJ, 676, 680

Fan, Y. 2009, Living Rev. Solar Phys., http://www.livingreviews.org/lrsp-2009-4

Fan, Y., Alexander, D., and Tian, L. 2009, ApJ, 707, 604

Hale, G. E., Ellerman, F., Nicholson, S. B., and Joy, A. H. 1919, ApJ, 49, 153

Jeong, H., & Chae, J. 2007, ApJ, 671, 1022

Kusano, K., Maeshiro, T., Yokoyama, T., & Sakurai, T. 2002, ApJ, 577, 501

Leka, K. D., Canfield, R. C., McClymont, A. N., van Driel-Gesztelyi, L. 1996, ApJ, 462, 547

Kusano, K., Maeshiro, T., Yokoyama, T., & Sakurai, T 2004, Astronomical Society of the Pacific Conference Series, 325, 175

LaBonte, B., Georgoulis, M., Rust, D. 2007, ApJ, 671, 955

Longcope, D. 2004, ApJ, 612, 1181

López Fuentes, M., Démoulin, P., Mandrini, C., van Driel-Gesztelyi, L. 2000, ApJ, 544, 540

López Fuentes, M., Démoulin, P., Mandrini, C., Pevtsov, A., van Driel-Gesztelyi, L. 2003, Sol. Phys., 397, 305

Nindos, A., Zhang, J., and Zhang, H. 2003, ApJ, 594, 1033

Pariat, E., Démoulin, P., & Berger, M. 2005, A&A, 439, 1191

Pariat, E., Nindos, A., Démoulin, P., & Berger, M. 2006, A&A, 452, 623

Schuck, P. 2005, ApJ, 632, 53L

Schuck, P. 2006, ApJ, 646, 1358

Schuck, P. 2008, ApJ, 683, 1134

Tian, L., Alexander, D., Liu, Y., & Yang, J. 2005, Sol. Phys., 229, 63

Tian, L., & Alexander, D. 2008, ApJ, 673, 532

Tian, L., & Alexander, D. 2009, ApJ, 695, 1012

van Driel-Gesztelyi, L. & Culhane, J. 2009, Space Sci. Rev., 144, 351

Welsch, B., Fisher, G., Abbett, W., & Régnier, S. 2004, ApJ, 610, 410

Welsch, B., Abbett, W., De Rose, M., Fisher, G. and et al. 2007, ApJ, 670, 434

Welsch, B., Li, Y., Schuck, P., & Fisher, G. 2009, ApJ, 705, 821

Zirin, H. 1988, *Astrophysics of the Sun (Cambridge: Cambridge Univ. Press)*, 307

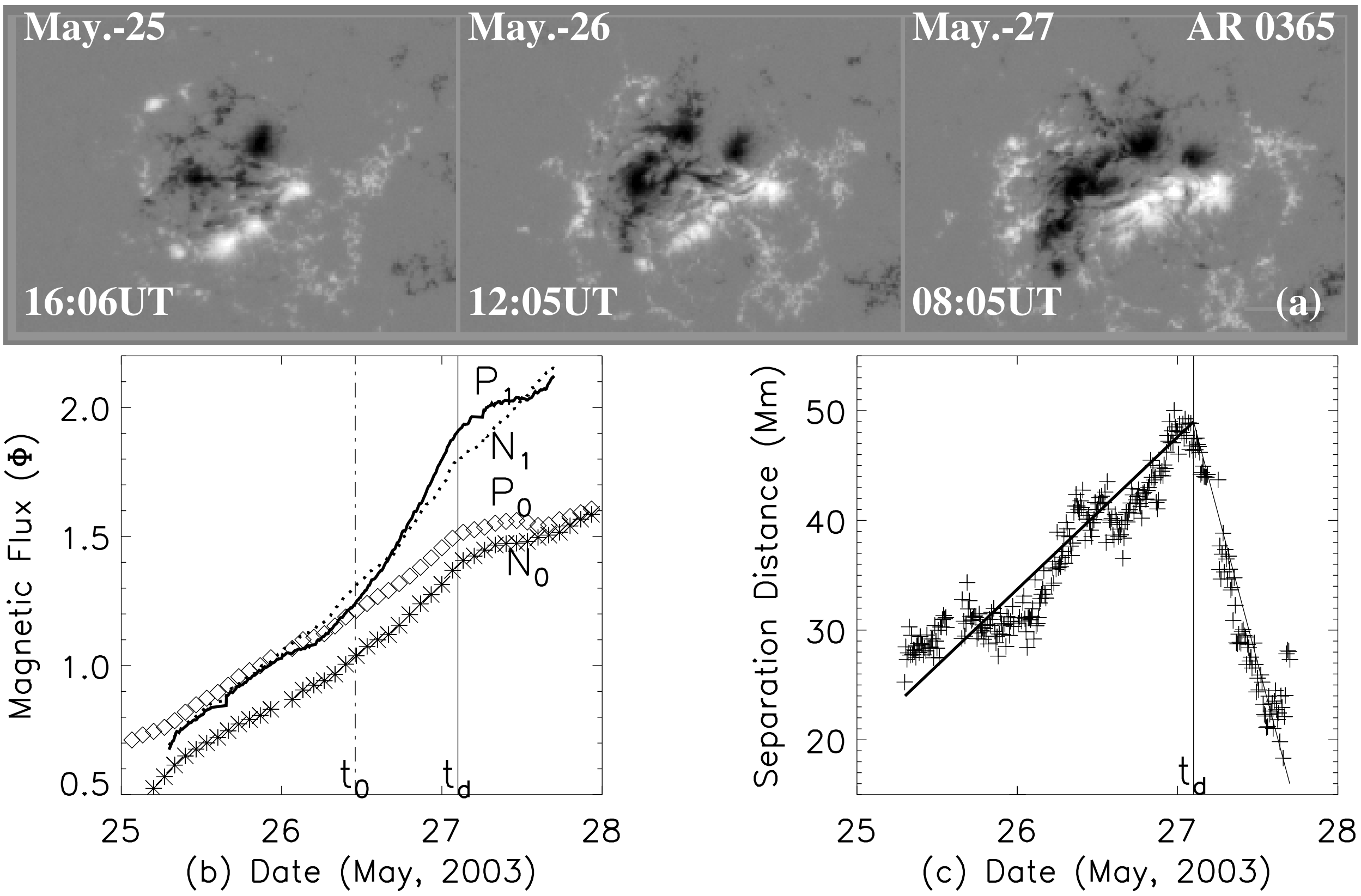

Fig. 1.— AR 10365 (S08): MDI/1m line-of-sight magnetograms with high spatial resolution (0.6 arcsecs) close to disk center (a), radial magnetic flux (b, unit of $10^{22}$ Mx) and polarity separation (c, unit of Mm). The size of the images is 210×150 arcsecs; black (white) patches denote negative (positive) magnetic field. '$\diamond$' (and $P_0$) and '$*$' (and $N_0$) denote positive and negative flux calculated by lev1.8.2 MDI/96m line-of-sight magnetograms (2.0 arcsecs/pixel). The bold curve (and $P_1$) and dotted curve (and $N_1$) denote 1.6 times the positive and negative flux calculated by lev1.5 MDI/1m magnetograms. Both magnetic fluxes and separation distances are computed with a threshold of 20 Gauss on the radial field component. The time '$t_0$' denotes the time when the AR was crossing the central meridian line, and the time '$t_d$' denotes the time when the polarity separation started to decrease.

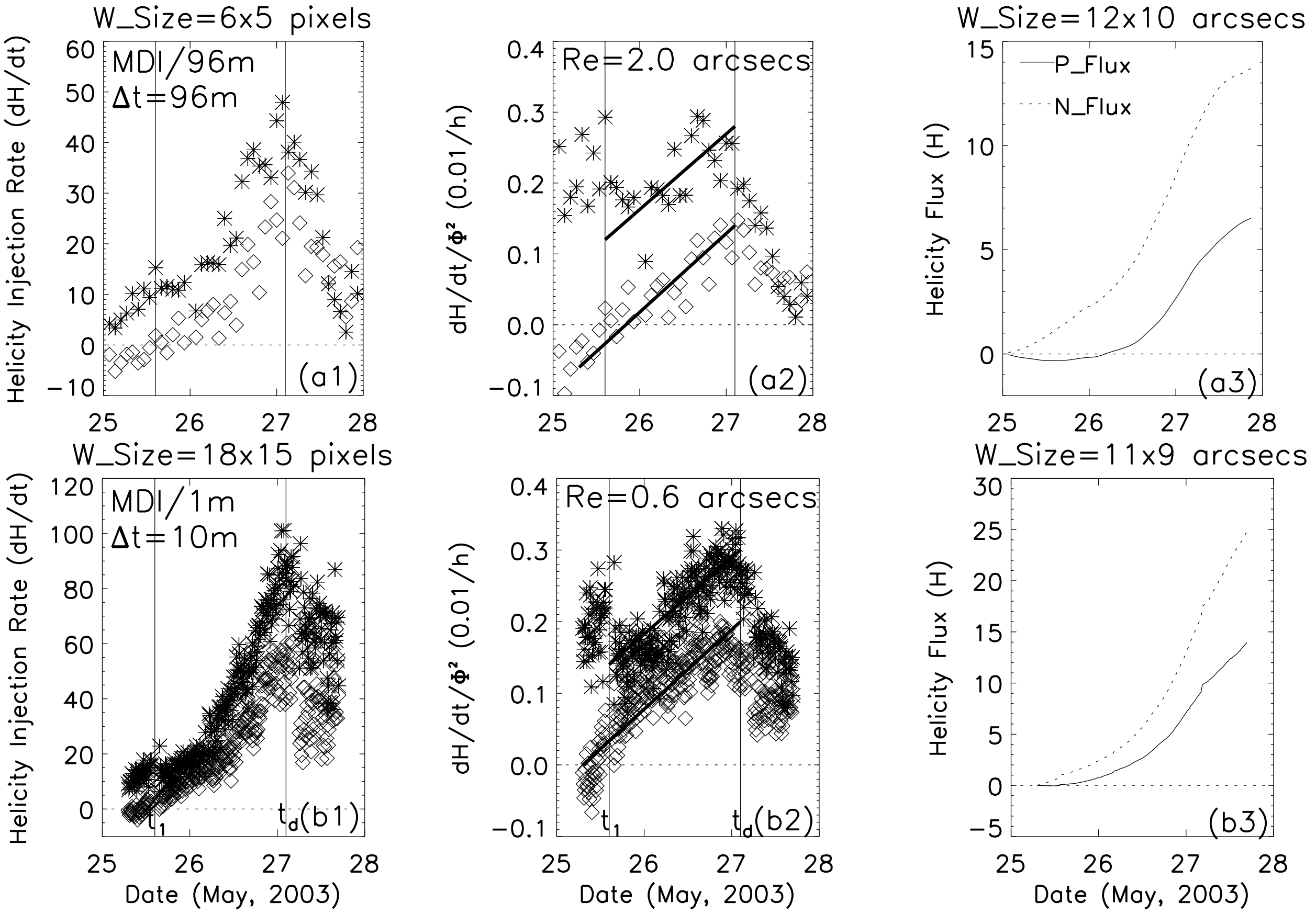

Fig. 2.— AR 10365: Magnetic helicity injection rate (dH/dt, unit of $10^{40}$ Mx$^2$/h), 'normalized' helicity injection rate (dH/dt/$\Phi^2$, unit of 0.01/h), and magnetic helicity flux (H=$\int(dH/dt)dt$, unit of $10^{42}$ Mx$^2$). (a1)−(a3) display results calculated from MDI/96m line-of-sight magnetograms with a window size of 12×10 arcseconds; (b1)−(b3) display results calculated from MDI/1m magnetograms with a window size of 11×9 arcseconds. '$\diamond$' is for the positive magnetic flux, while '$*$' is for the negative flux. The thick bold lines denote a linear trend of the evolution of the 'normalized' helicity rate between two times ($t_1$ and $t_d$).

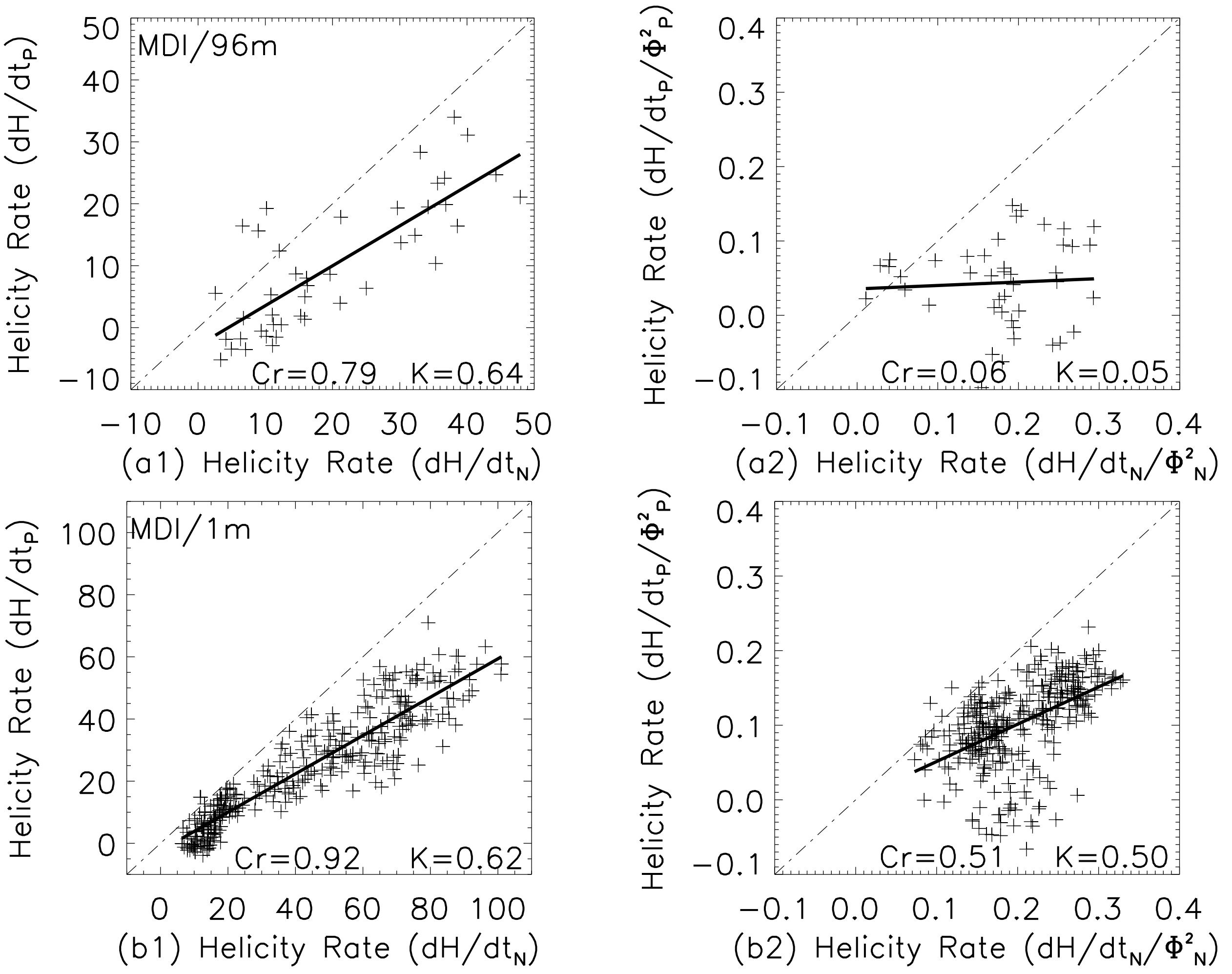

Fig. 3.— AR 10365: Relationship of magnetic helicity rate injected by positive magnetic flux ($dH/dt_P$ and $dH/dt_P/\Phi_P^2$) and negative magnetic flux ($dH/dt_N$ and $dH/dt_N/\Phi_N^2$). (a1) and (a2) are from MDI/96m data, while (b1) and (b2) are from MDI/1m data. 'K' and 'Cr' are slopes of linear fits and the correlation coefficients. Same window size and time interval to that shown in Figure 2.

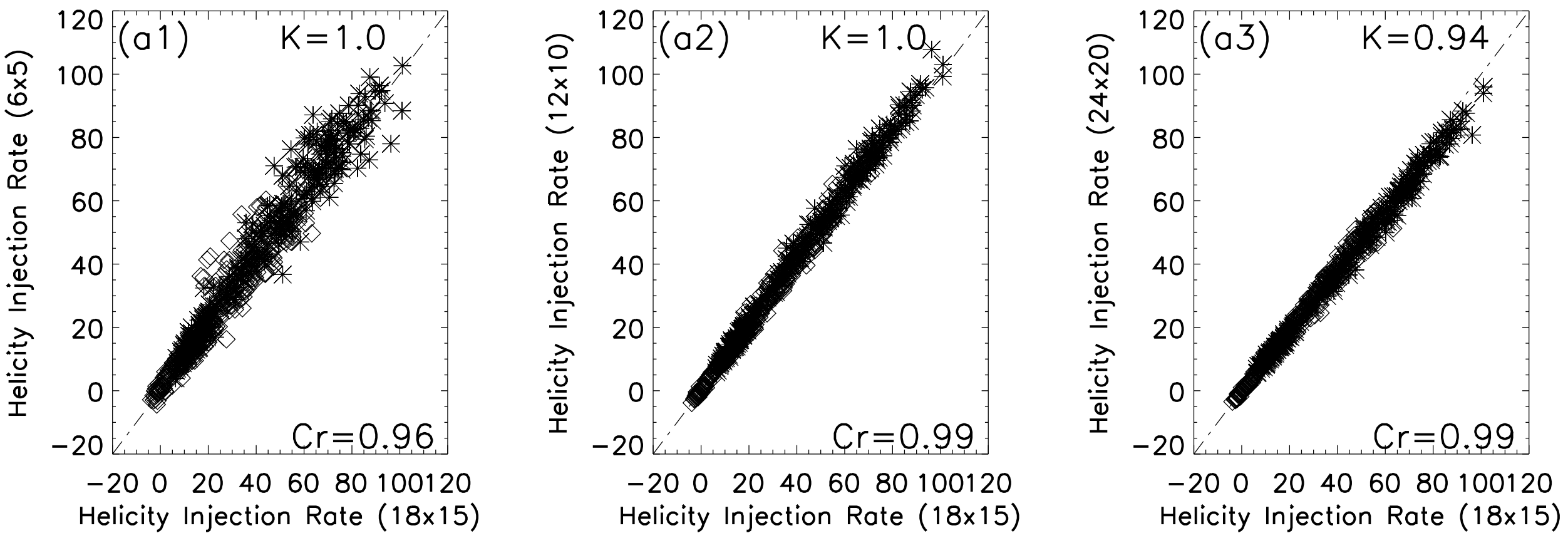

Fig. 4.— AR 10365: y-axis denotes the magnetic helicity injection rate at window size=$6 \times 5$, $12 \times 10$, and $24 \times 20$ pixels; x-axis denotes the injection rate at window size=$18 \times 15$ pixels. The time interval $\Delta t$ is 10 minutes. 'K' and 'Cr' are slopes of linear fits and the correlation coefficients. $\diamond / *$ denotes magnetic helicity rate injected by the positive/negative magnetic flux.

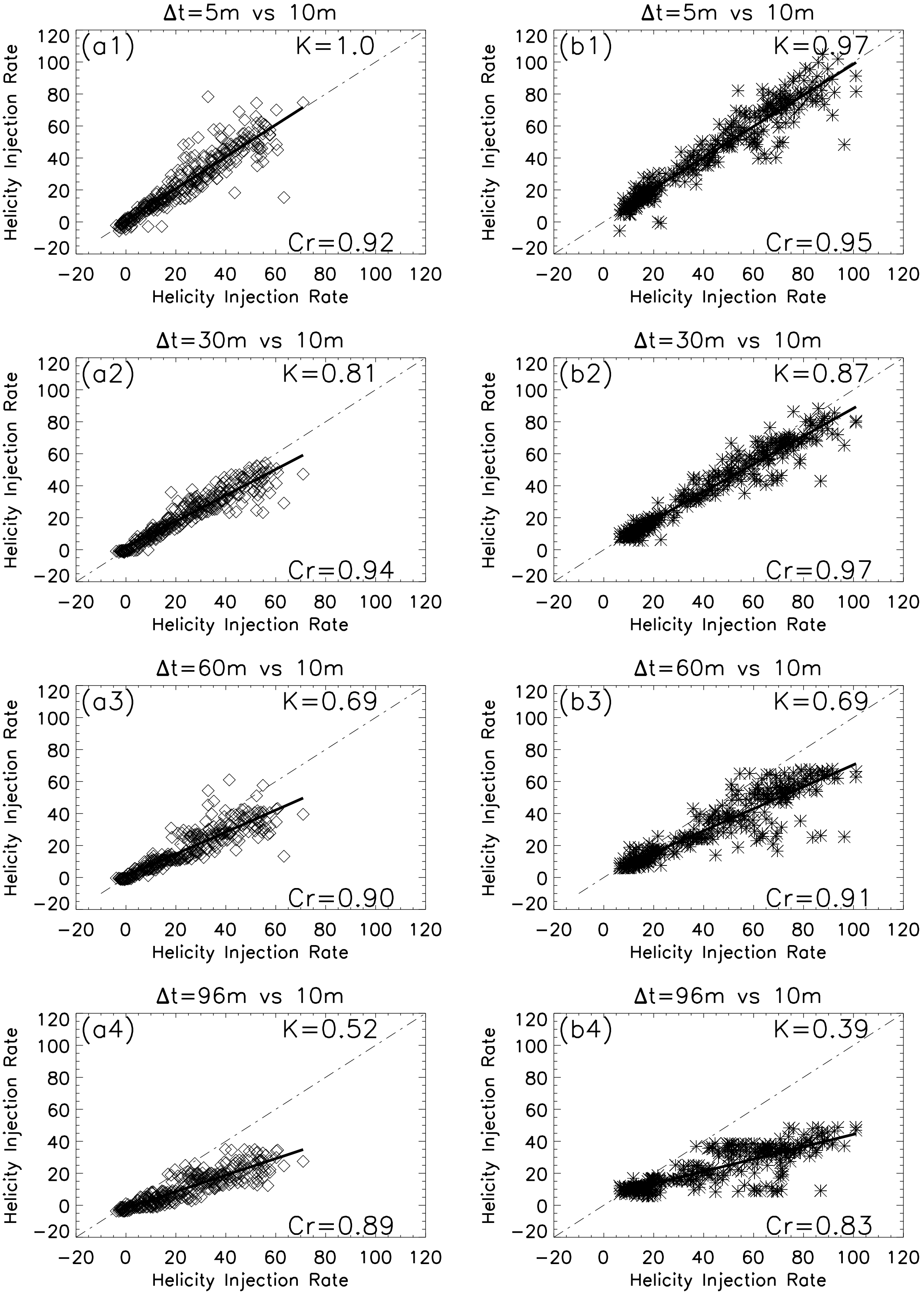

Fig. 5.— AR 10365: y-axis denotes a cubic spline interpolation for the magnetic helicity injection rate at $\Delta t$=5m, 30m, 60m, and 96m; x-axis denotes the injection rate at $\Delta t$=10m. The window size is 18 × 15 pixels. NOTE: (a4) and (b4) are from MDI/96m data, while others are from MDI/1m data. 'K' and 'Cr' are slopes of linear fits and the correlation coefficients. $\diamond/*$ denotes the rate injected by the positive/negative magnetic flux.

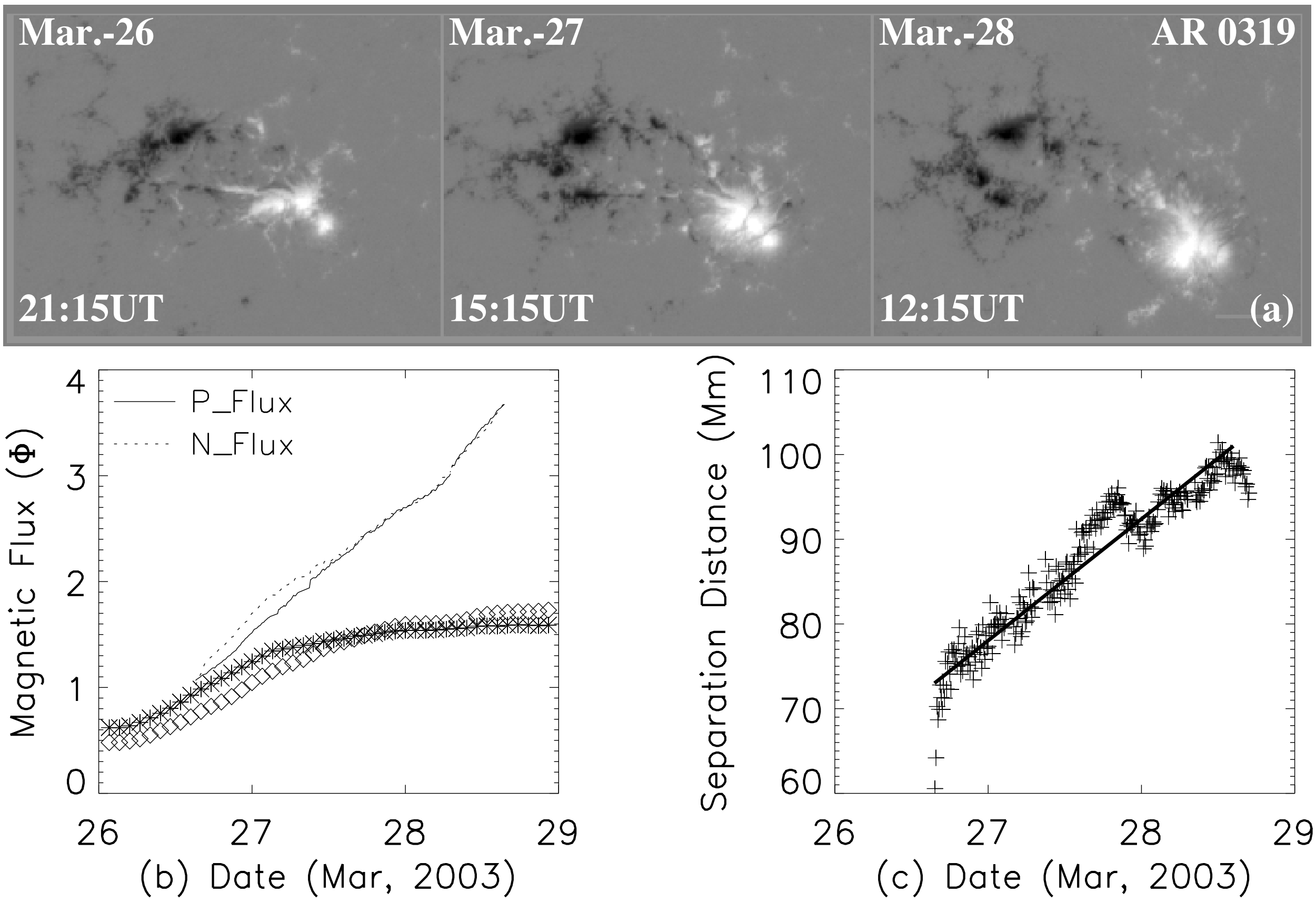

Fig. 6.— MDI/1m line-of-sight magnetograms for AR 10319 (N13). Size of the images is 240×180 arcseconds. Others are same to Figure 1.

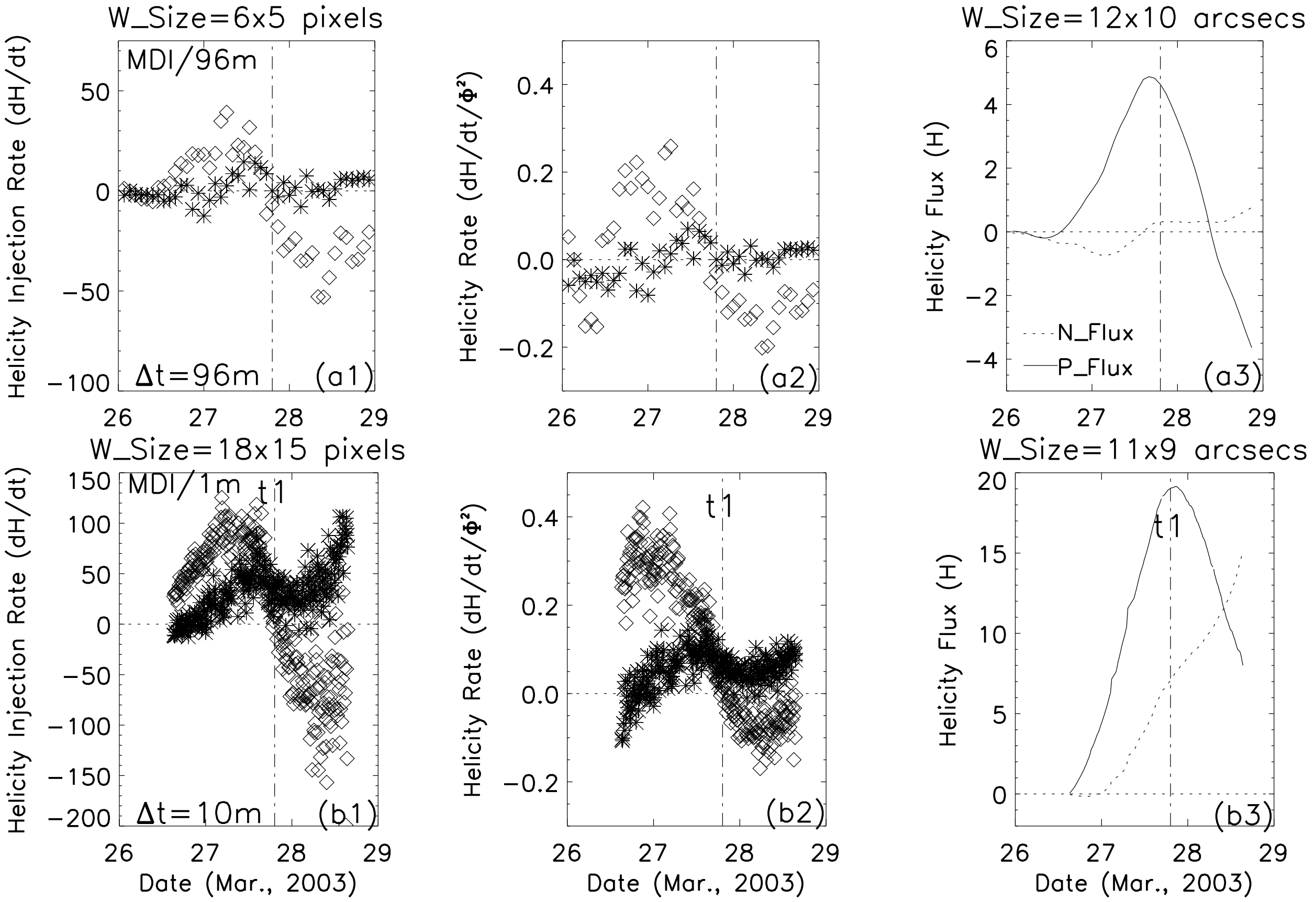

Fig. 7.— Magnetic helicity injection for AR 10319. a(1−3) is for MDI/96m data, while b(1−3) is for MDI/1m data. '◇' and '∗' denote the magnetic helicity rates injected by positive and negative magnetic flux. Dashed-dotted vertical lines denote the time when the active region was crossing the central meridian line. Others are same to Figure 2.

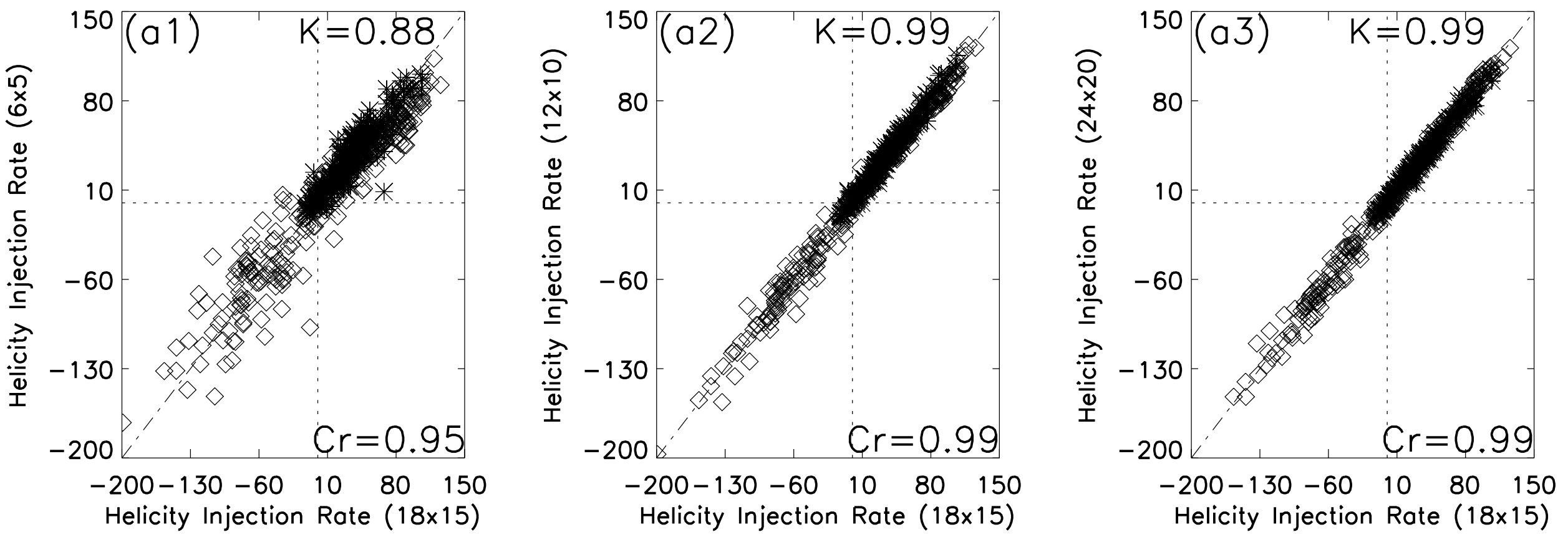

Fig. 8.— Magnetic helicity injection rate for AR 10319. 'K' and 'Cr' are slopes of linear fits and the correlation coefficients. Others are same to Figure 4.

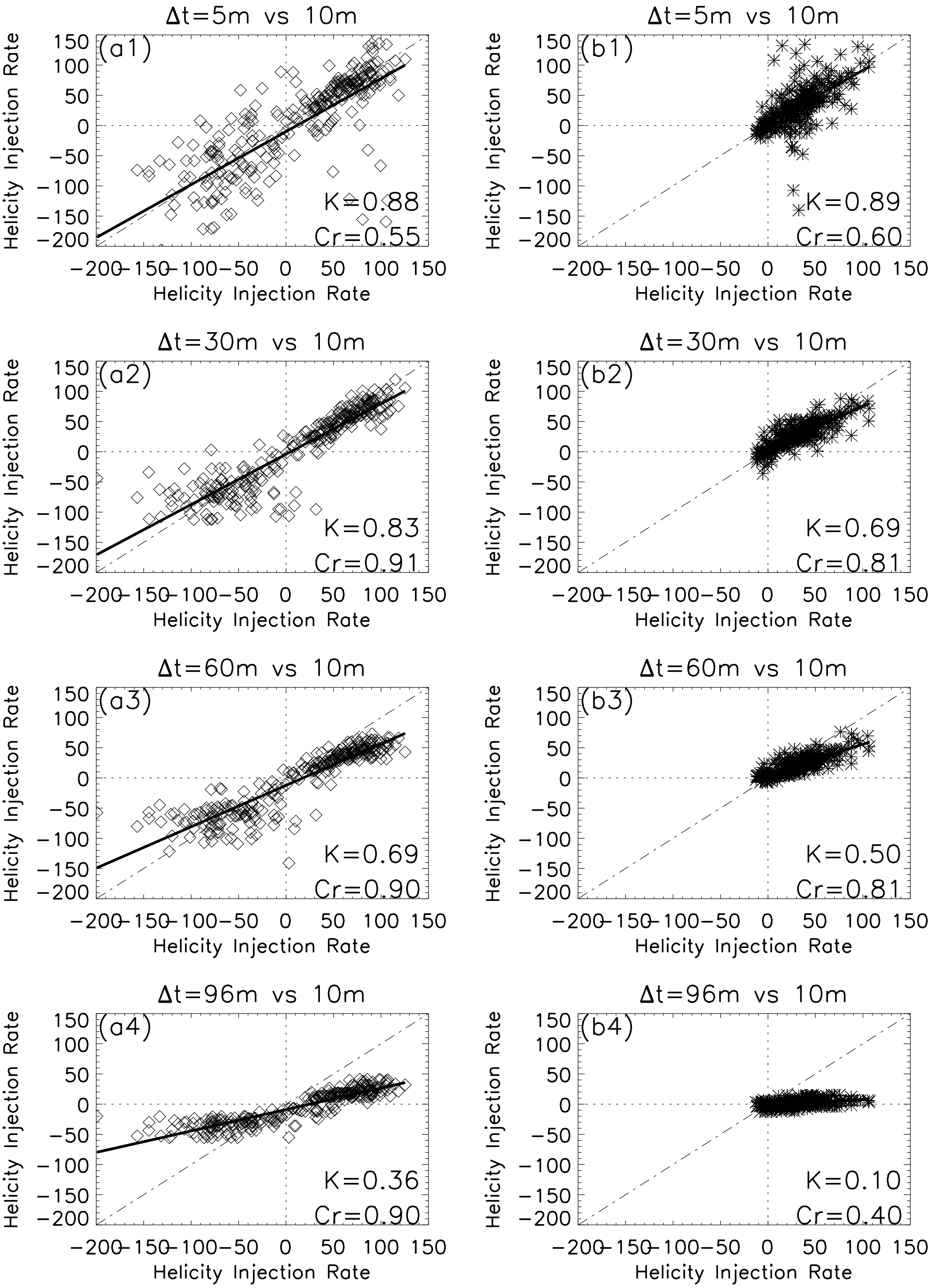

Fig. 9.— Magnetic helicity injection rate for AR 10319. ‘K’ and ‘Cr’ are slopes of linear fits and the correlation coefficients. Others are same to Figure 5.

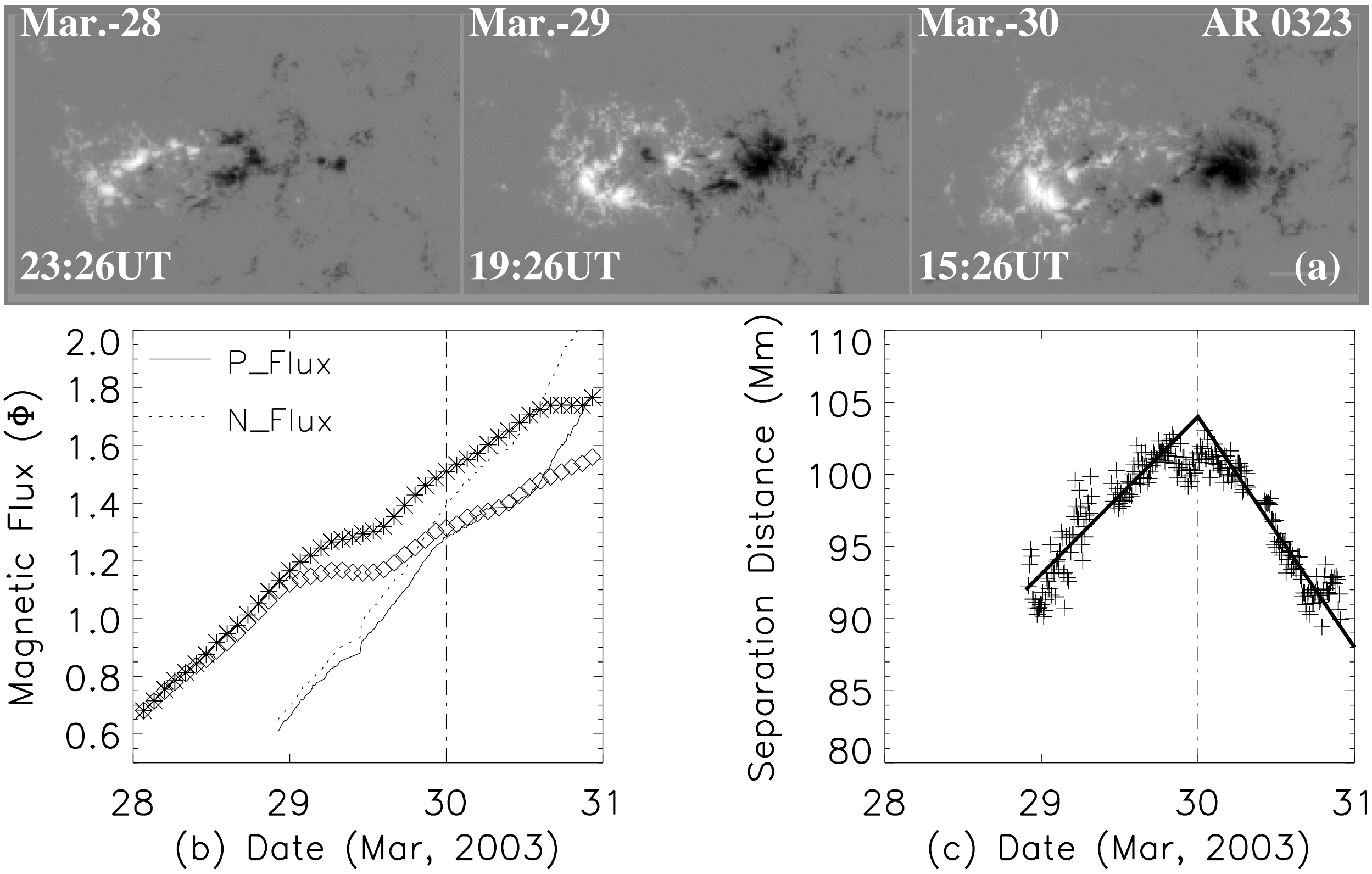

Fig. 10.— MDI/1m line-of-sight magnetograms for AR 10323 (S09). Size of the images is 240×180 arcseconds. Dashed-dotted vertical lines denote the time when the active region was crossing the central meridian line. Others are same to Figure 1.

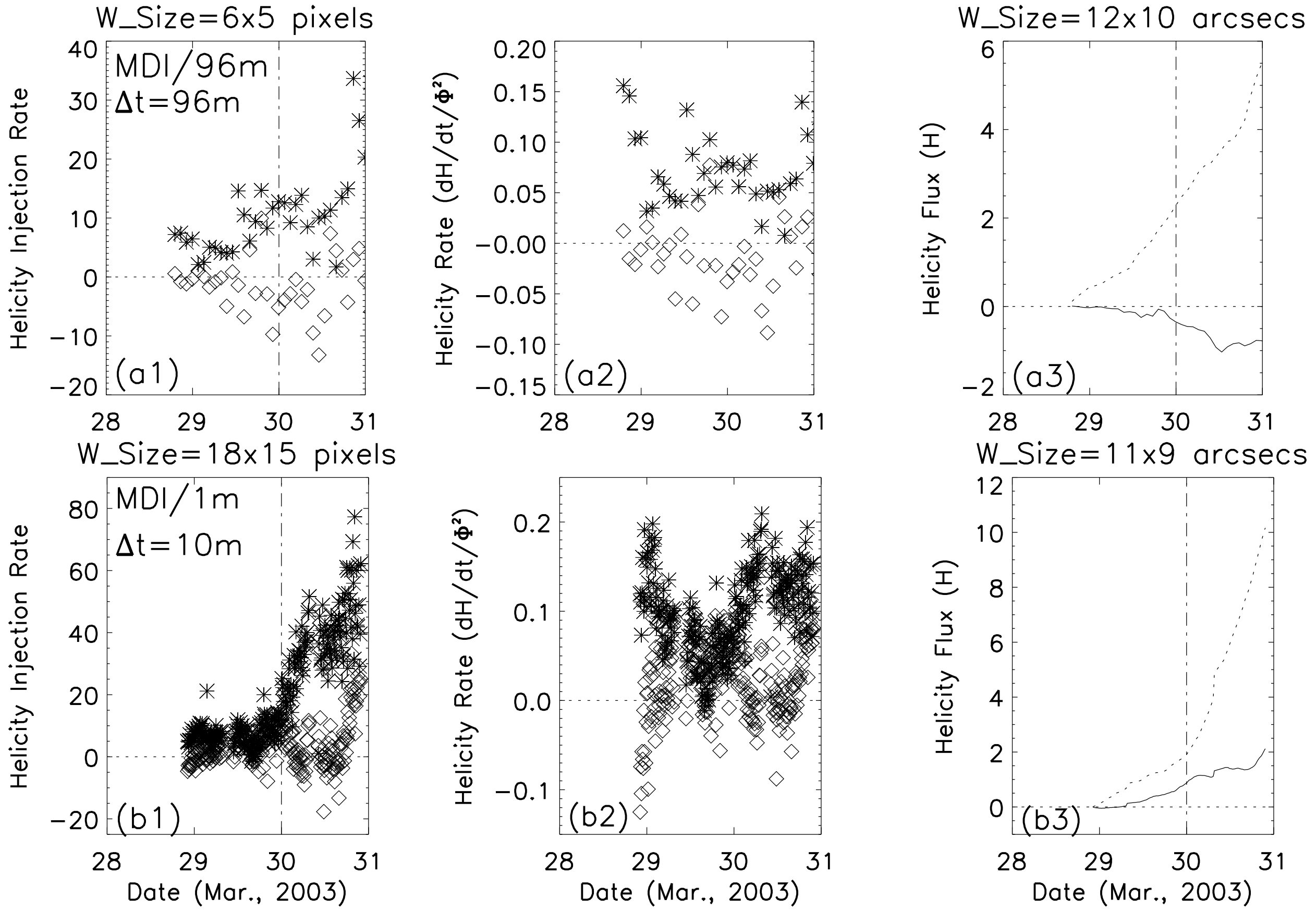

Fig. 11.— Magnetic helicity injection for AR 10323. Dashed-dotted vertical lines denote the time when the active region was crossing the central meridian line. Others are same to Figure 2.

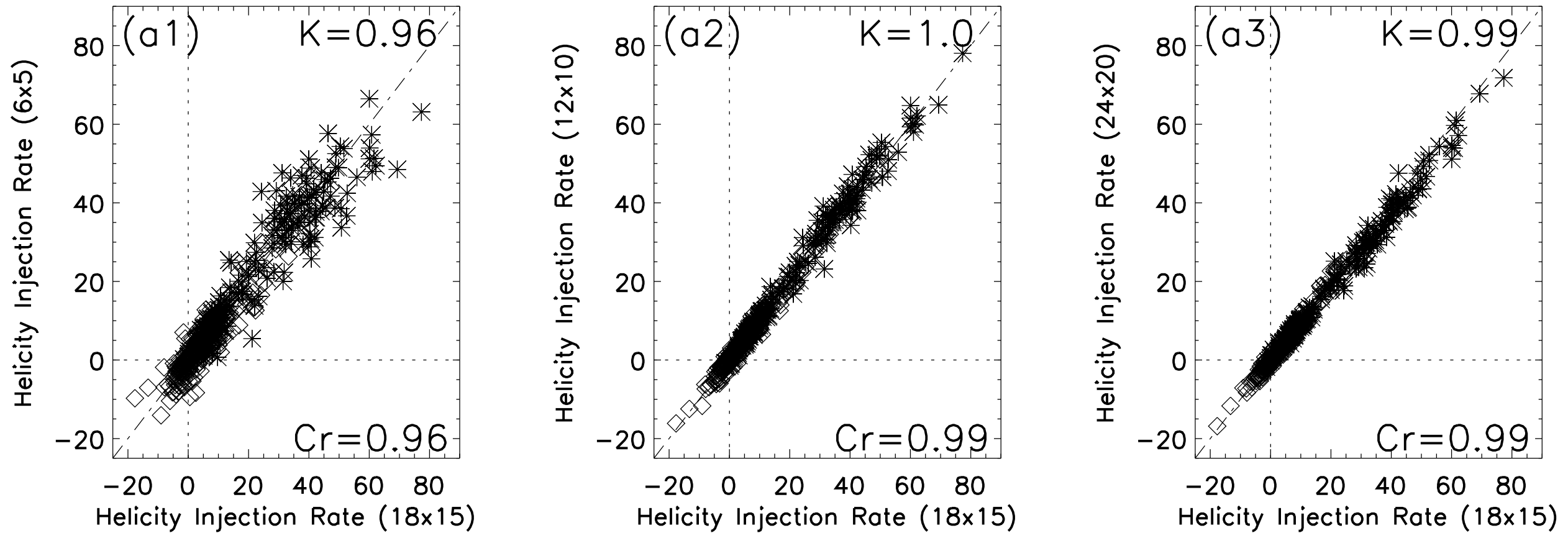

Fig. 12.— Magnetic helicity injection rate for AR 10323. ‘K’ and ‘Cr’ are slopes of linear fits and the correlation coefficients. Others are same to Figure 4.

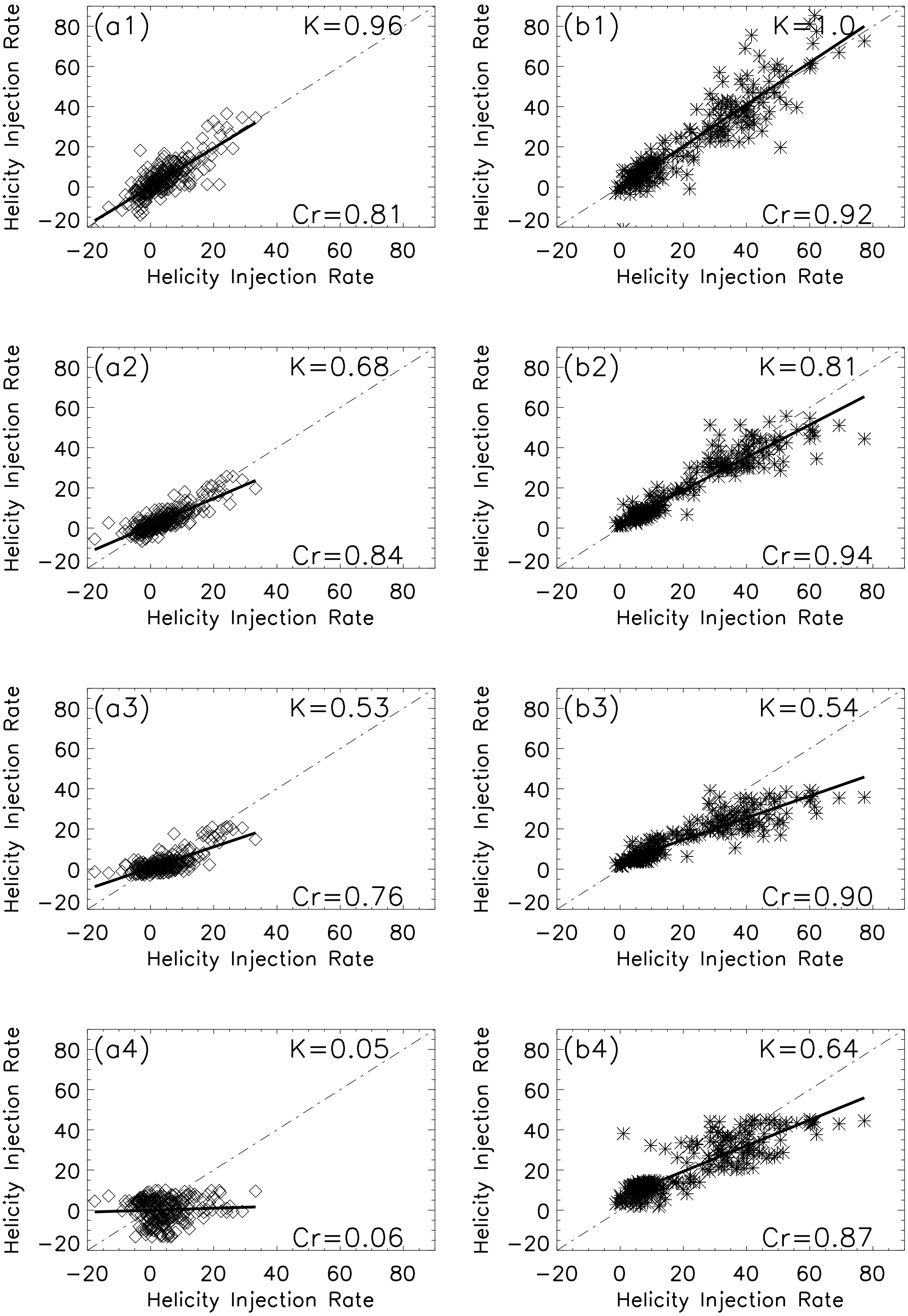

Fig. 13.— Magnetic helicity injection rate for AR 10323. 'K' and 'Cr' are slopes of linear fits and the correlation coefficients. Others are same to Figure 5.

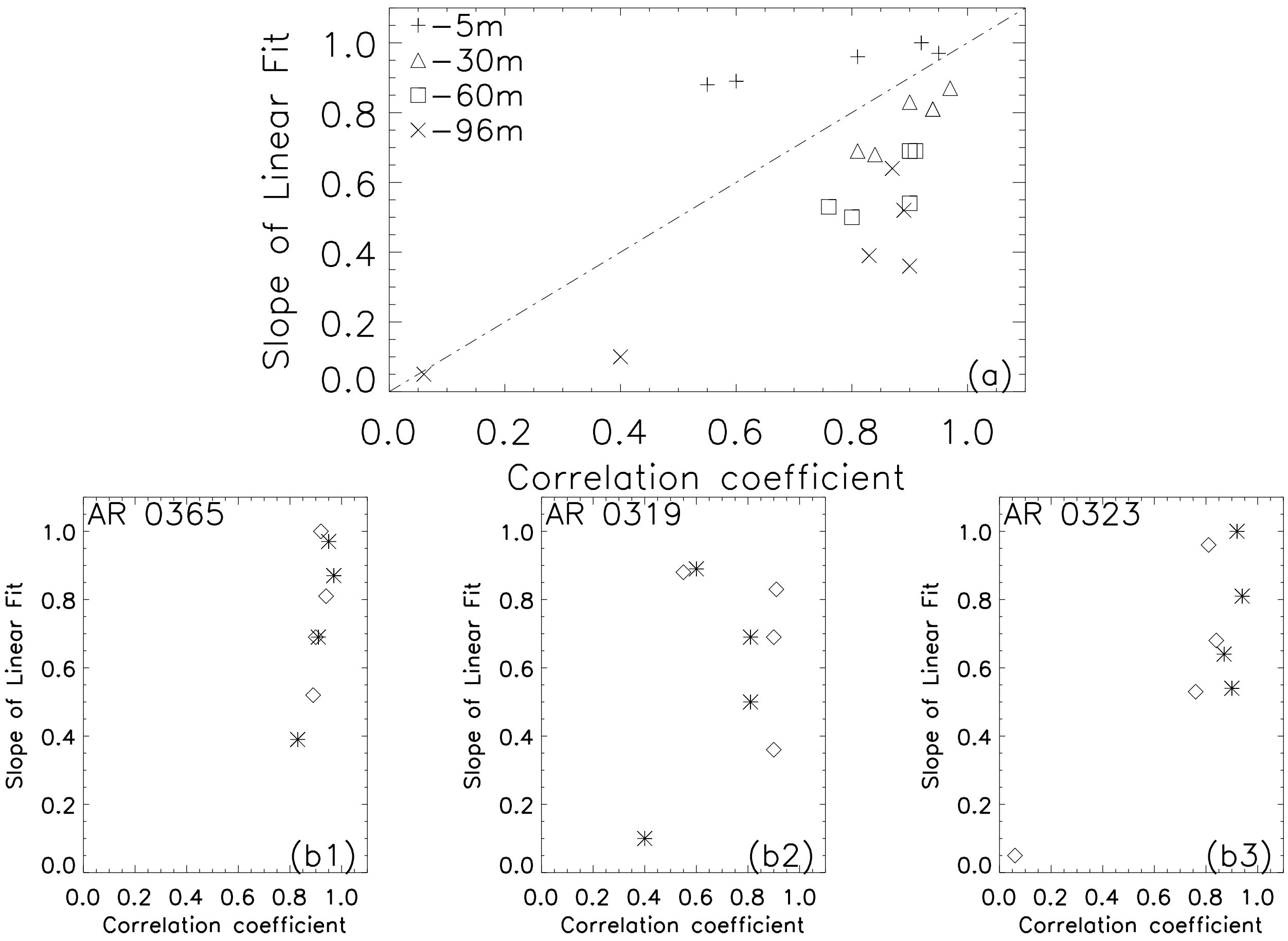

Fig. 14.— Relationship of slopes ($K$) of a linear fit and correlation coefficients (Cr), shown in Figs.-5, 9, and 13. (a): for two polarities of all three active regions at different time interval $\Delta t$; (b): for each active region, where $\diamond/*$ denotes the value related to the positive/negative magnetic flux, following/leading polarity for ARs 10365 and 10323, while leading/following polarity for AR 10319 (see also Table 1).

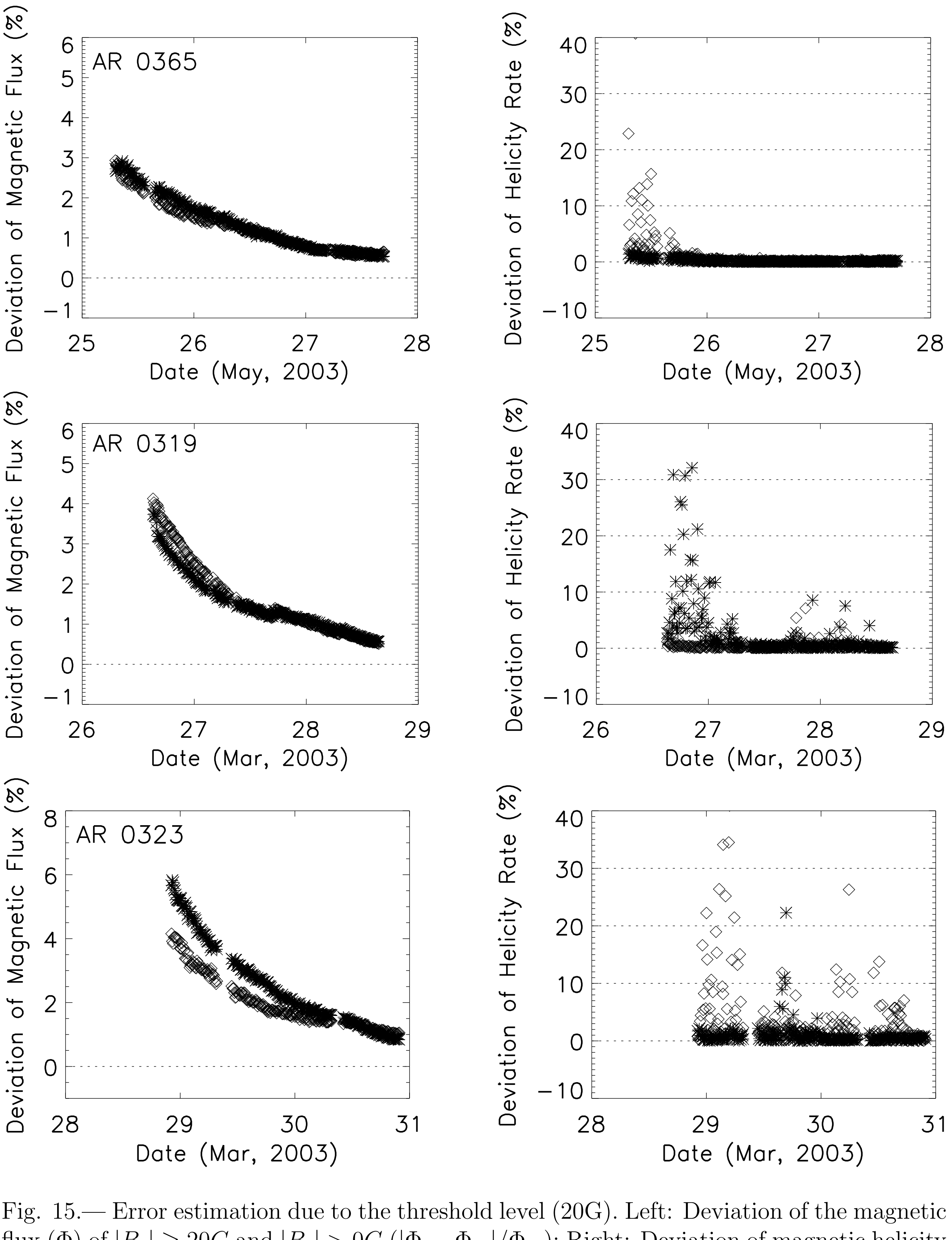

Fig. 15.— Error estimation due to the threshold level (20G). Left: Deviation of the magnetic flux ($\Phi$) of $|B_r| \geq 20G$ and $|B_r| > 0G$ ($|\Phi_0 - \Phi_{20}|/\Phi_{20}$); Right: Deviation of magnetic helicity injection rate (dH/dt) over $|B| \geq 20G$ and $|B| > 0G$ ($|dH/dt_0 - dH/dt_{20}|/(dH/dt_{20})$).